\documentclass[preprint,12pt]{aastex}
\usepackage{epsfig}
\usepackage{emulateapj5,apjfonts}
\begin{document}

\def\wdf{white dwarf}
\def\etal{et al.} 
\def\rd{Di\thinspace Stefano}
\newcommand{\chandra}{{\it Chandra}}
\newcommand{\asca}{{\it ASCA}}
\newcommand{\rosat}{{\it ROSAT}}
\newcommand{\sax}{{\it BeppoSAX}}
\newcommand{\xmm}{{\it XMM}}
\newcommand{\einstein}{{\it Einstein}}
\newcommand{\lum}{\thinspace\hbox{$\hbox{erg}\thinspace\hbox{s}^{-1}$}}

\def\spose#1{\hbox to 0pt{#1\hss}}
\def\laeq{\mathrel{\spose{\lower 3pt\hbox{$\mathchar"218$}}
     \raise 2.0pt\hbox{$\mathchar"13C$}}}
\def\gaeq{\mathrel{\spose{\lower 3pt\hbox{$\mathchar"218$}}
     \raise 2.0pt\hbox{$\mathchar"13E$}}}

\title{The Discovery of Quasisoft and Supersoft Sources in External Galaxies}

\author{R.~Di\,Stefano$^{1,2}$, A.~K.~H.~Kong$^{1}$} 
\affil{$^1$ Harvard-Smithsonian Center for Astrophysics, 60
Garden Street, Cambridge, MA 02138}
\affil{$^2$ Department of Physics and Astronomy, Tufts
University, Medford, MA 02155}

\begin{abstract}

We apply a uniform procedure to select very soft sources from point sources observed by {\it Chandra} in
$4$ galaxies. This sample
includes one elliptical galaxy (NGC 4967), $2$ face-on spirals
(M101 and M83), and an interacting galaxy (M51). 
We report on some intriguing results, including the following.

\noindent (1) We have found very soft X-ray sources (VSSs) in every galaxy. Some of these fit the criteria for canonical supersoft sources (SSSs), while others are somewhat harder. These latter have characteristic values of $kT \laeq 300$~eV; we refer to them as quasisoft sources (QSSs). We found a combined total of $149$ VSSs in the $4$ galaxies we considered; 77 were SSSs and 72 were QSSs.

\noindent (2) The data are consistent
with the existence of a large VSS population, most of whose members
we cannot observe
due to the effects of distance and obscuration. The total VSS population 
of sources with $L > 10^{37}$ erg s$^{-1}$  
in each galaxy could be on the order of $1000$. 

\noindent (3)  
Whereas in M31 only $\sim 10\%$
of all X-ray sources detected by {\it Chandra} are VSSs,  
more than $35\%$ 
of all detectable X-ray sources in the face-on galaxy M101 
fit the phenomenological definition of VSSs. 
This difference may be due to
differences in $N_H$ between typical lines of sight to sources in
each galaxy. 

\noindent (4) SSSs can be super-Eddington for Chandrasekhar mass objects.

\noindent (5) We find evidence for SSSs and QSSs with
luminosities $10^{36}$ erg s$^{-1} < L < 10^{37}$ erg s$^{-1}$.
These sources have luminosities lower than those of the $\sim 30$ 
soft sources used to establish the class of SSSs.

\noindent (6) In   
the spiral galaxies M101, M83 and M51, 
a large fraction of the SSSs and QSSs appear to
be associated with the spiral arms. This may indicate that
some SSSs are young systems, possibly younger than $10^8$ years.

\noindent (7) In addition to finding hot white dwarfs
and soft X-ray binaries, our method should also be efficient
at selecting supernova remnants (SNRs).
A small fraction of the VSSs in the spiral 
arms of M101 appear to be associated with SNRs.

\end{abstract}

\keywords{black hole physics --- galaxies: individual (M101, M83, M51, NGC 4697) --- supernova remnants --- white dwarfs --- X-rays: binaries --- X-rays: galaxies}

\section{Introduction}

\chandra\ and \xmm\ observations make it
possible to systematically discover and study very soft X-ray sources (VSSs)
in external galaxies. Until now, the canonical examples of VSSs were 
luminous supersoft X-ray sources (SSS). The SSSs that established the class lie in the Magellanic Clouds and in the Milky Way; they have $kT$ in the range of
tens of eV
and luminosities
 between roughly $10^{37}$ erg s$^{-1}$ and a few times $10^{38}$
erg s$^{-1}$. \chandra\ and \xmm\ surveys in other galaxies are important, since gas obscures 
more than 99\% of the Milky Way's SSSs, so that direct studies of the size and
characterics of galactic populations of SSSs can be carried out only
in other galaxies. 

This paper is the third is a series on the selection of VSSs in external
galaxies. In paper 1 (Di Stefano \& Kong 2003a), we presented a set of
strict hardness ratio conditions (the ``HR conditions) to select only
the softest sources. Sources selected by the HR conditions are
likely to have spectra very similar to those of the SSSs
observed in the Galaxy and Magellanic Clouds. Also in paper 1
we tested the HR conditions by applying them to {\it Chandra} data
from $4$ galaxies. In paper 2 (Di Stefano \& Kong 2003b), we presented an
algorithm that starts with the HR conditions, but which then proceeds to
progressively relax them. Each set of relaxed conditions chooses sources
that deviate from the expected broadband spectra of SSSs in some way.
For example, the $3\sigma$ conditions select sources that may be as soft
as SSS-HRs, but which provide too few photons to satisfy the
stronger HR conditions. Any source satisfying either the HR or
$3\sigma$ conditions is called a
classical supersoft source, or simply an SSS (SSS-HR or SSS-$3\sigma$). 
In addition,
there is a set of $7$ conditions which can select either
highly absorbed SSSs, or sources that are genuinely harder than
the SSSs studied in the Galaxy and Magellanic Clouds. We refer to
any source which satisfies one of these weaker conditions
as a quasisoft source (QSS). For example, the
``NOH'' conditions
can select quasisoft sources (QSS-NOH) located behind large gas columns.
Other conditions (e.g., the ``$\sigma$'' conditions) select sources 
(QSS-$\sigma$)
which may have a dominant component as soft as SSSs, but
which also include a small hard component. All sources selected
by the algorithm, whether they are supersoft or quasisoft,
are referred to as very soft sources (VSSs). 

In this paper, we apply the full algorithm to data from the same
$4$ galaxies considered in paper 1: M101, M83, M51, and NGC 4472.
Because the HR conditions have already been applied, we could
have left SSS-HR sources out of this analysis. We decided however,
to include them, because it is useful to explicitly compare their
properties with those of the other soft sources identified
by higher steps in the algorithm.

\section{Data Analysis}

For M101 and M51, we generated our own source lists with the CIAO tool
WAVDETECT. For M83 we used a source list provided Soria \& Wu (2003), and for
NGC 4697 we used the source list of Sarazin et al. (2001). For M101 and M51, we set the input parameters for WAVDETECT to cover the energy range 0.1--7 keV. For galaxies in which the lower energy limit used by WAVDETECT was greater than 0.1 keV, we carried out a visual inspection to search for any sources with
reliable detections but with few photons detected above 0.3 keV.
Source counts in each energy band were
determined via aperture photometry. The radius of the aperture was
varied with off-axis angle in order to match the 90\% encircled energy
function. Background was chosen from an annulus region centered on each
source. We note that, while the calibration at the lowest
energies ($k\, T < 0.3$ keV) will need to be refined 
for spectral fits, source detection algorithms should include
photons with energies as low as $0.1$ keV if all SSSs and QSSs are
to be detected. In $3$ of the galaxies, photons with energies
between $0.1$ keV and $0.3$ keV have helped to detect a small number of SSSs and QSSs.
Because visual inspection of each source must be carried out
anyway, any spurious sources detected because of background
effects at low energy 
can be easily eliminated.

In addition to estimates of the number of background objects based on
the {\it Chandra} Deep
Field data (see the references in each of the galaxies below), we also used the ChaMP ({\it Chandra}
Multiwavelength Project \footnote{http://hea-www.harvard.edu/CHAMP})
archives (P. Green, private communication) to estimate the contamination from soft X-ray sources in the foreground (X-ray active stars) and background (e.g., soft AGN).
In 5 ACIS-S observations
with durations of $\sim 10-20$ ks, only 3 sources were found to be QSSs; no SSSs were identified. We therefore believe that the
background contribution of soft sources is small compared with the
large population of SSSs in our sample galaxies.

\subsection{The Sources}

\subsubsection{M101}

M101 (= NGC5457) is a face-on nearby (5.4--6.7 Mpc) spiral and was 
observed by {\it
Chandra} ACIS-S for
about 100ks on 2000 March 26-27. Detailed results of the observations have
been reported by Pence et al. (2001). We generated our own source list
here with CIAO task WAVDETECT instead of CELLDETECT used by Pence et al.
(2001). We also used different energy bands covering 0.1--7 keV,
instead of 0.125--8 keV in Pence et al. (2001). Pence et al. (2001)
estimated that 27 sources are possible background AGN.
While the source lists agree with each other generally, we found
8 additional sources (118 in total) in the S3 chip. The list of
SSSs and QSSs is shown in Table 2. We found a total of 53 VSSs; 32 were SSSs and 21 were QSSs.

\subsubsection{M83}

M83 (= NGC5236) is a barred spiral galaxy with low inclination angle ($i=24^{\circ}$); 
it has a starburst nucleus and the estimated distance ranges from 3.7 Mpc 
to 8.9 Mpc. A {\it Chandra} ACIS-S observation was taken on 2000 April 29
for about
50ks. The X-ray point source properties and the nuclear region were discussed 
by Soria and Wu (2002,2003). One hundred and twenty seven point sources were found in the S3 chip (0.3--8 keV) and approximately 10 sources are likely to be
background AGN. We used this 
source list for our analysis. In addition, visual inspection from a
0.1--1 keV image discovered 1 new SSS. Table 3 lists M83's SSSs and QSSs. We found a total of 54 VSSs; 28 were SSSs and 26 were QSSs.

\subsubsection{M51}

M51 (= NGC5194) is a nearby (7.7--8.4 Mpc) interacting spiral galaxy with
moderate inclination ($i=46^{\circ}$). A {\it Chandra} ACIS-S observation was
performed on 2000 June 20 for about 15ks. A detailed analysis of the data is
given by Terashima \& Wilson (2003). We generated our own source list with the CIAO
task WAVDETECT. A total of 72 sources (in 0.3--7 keV) were found in the
S3 chip. About 10 of the sources are expected to
be background objects. All SSSs and QSSs are listed in Table 4. We found a total of 23 VSSs; 13 were SSSs and 10 were QSSs.

\subsubsection{NGC4697}

NGC4697 is an elliptical galaxy at a distance of 11.7--23.3 Mpc. It was observed
by {\it Chandra} ACIS-S
on 2000 January 15-16 for about 40ks. Ninety point sources were detected in the
S3 chip (0.3--10 keV; Sarazin, Irwin, \& Bregman 2001); about 10--15
background objects are expected in the observation. The complete source list
and results from this observation can be found in Sarazin et
al. (2001). We also inspected an image from 0.1--1 keV and found 1 new
uncatelogued SSS.
Table 5 lists the SSSs and QSSs found with our selection procedure. We found a total of 19 VSSs; 4 were SSSs and 15 were QSSs.

\subsection{Spectra of Very Soft Sources}

\subsubsection{The Spectra}

\noindent {\sl Individual Sources:\, }  
To determine whether our algorithm selects sources that 
are genuinely soft, we have extracted the energy spectra
of all very soft sources that provided more than $\sim 200$ counts. 	
Background counts were extracted in annulus regions centered on each 
source.
There are $10$ such sources.
We have also considered the spectra of sources with
more than $100$ counts, although the uncertainties in spectral parameters   
are large. A representative spectrum of a QSS-$\sigma$ source is shown in Figure 1. The spectral fits are list in Table 6.

\noindent {\sl Composite Spectra:\, }
Because many VSSs are faint, fits of individual spectra are not  
possible. If, therefore, we found a number of sources
that (1) had been identified by the same criteria, and which
(2) displayed similar broadband spectra 
and (3) provided a total of approximately $100$ counts,
we extracted
the composite energy spectra (see e.g., Figure 1). 

\noindent Because the response matrices vary across the detector, a weighted
RMF and ARF was generated following the thread recommended by the {\it
Chandra} X-ray Center\footnote{http://asc.harvard.edu/ciao/threads/wresp\_multiple\_sources}.
Results are presented together with the individual spectra in Table 6.

The analysis of {\it Chandra} spectra of VSSs is fraught with
uncertainties
because the low-energy ($k\, T < 0.5$ keV)
calibration of ACIS-S  is not yet well understood \footnote{
We have, e.g., data on all of the nearby SSSs, and find anomalous
behavior at low energies that appears to be related to the difficulties
encountered so far with the low-energy calibration.
Some of these anomalies would be less obvious in the distant
sources we discuss here, because the total number of counts is
generally relatively small; calibration uncertainties
 may nevertheless affect the 
derived values of spectral parameters. 
}.
It is therefore important to keep in mind that
the spectral fits of SSSs, especially those that peak at   
energies near or below $0.5$ keV are not on the same footing as 
the spectral fits of higher-energy sources, and may need to
be reanalyzed once the low-energy calibration has been
better studied.  

We also note that the quantum
efficiency degradation of ACIS at low energies 
is important for SSSs and QSSs.  
The
degradation has been
 shown to be a function of time and is most severe at low
energies\footnote{see 
http://asc.harvard.edu/cal/Links/Acis/acis/Cal\_prods/qeDeg/}. For a
source with blackbody temperature of 100 eV and 
$N_H=10^{20}$ cm$^{-2}$, the ACIS-S counts below 2 keV had dropped by about 30\%
one year after the launch and by almost 50\% after two years (P. Plucinsky,
private communication). 
Fortunately this problem has been well studied and it is possible to
take the effect into account.
We therefore
applied corrections using the ACISABS absorption model in XSPEC
\footnote{http://www.astro.psu.edu/users/chartas/xcontdir/xcont.html}.
This model allows us to correct the response by inputting the number of
days between {\it Chandra's} launch and the observation. We find that, for very soft sources, the degradation mainly affects the best 
fit value of $N_H;$ the
uncorrected model tends to over-estimate the value (see also Kong et
al. 2002). 
 
For each spectral fit, we selected photons
within a  circular region centered on the source. Only $0.3-7$ keV data were
used. We first fit the spectra with simple one-component models
including absorbed power-law, blackbody and Raymond-Smith (RS) models.
In all but three cases, an absorbed blackbody model can provide a good fit.
For the three exceptional spectra, we added an additional power-law component
or fit with a RS model in order to derive an acceptable result.

Table 6 shows that, as expected, sources satisfying the HR conditions are generally the softest. While the other sources tend to be harder, all but 1 satify the conditions used to define VSSs (Di\,Stefano \& Kong 2003a); M51-10 appears to be marginally harder. 
We note one unusual feature. This is that, in each galaxy,
 typical values of $N_H$ associated with the fits of QSSs are 
significantly higher than the minimum value of $N_H$ along the  
direction to the galaxy. If, however,
 internal absorption is not a significant factor for most SSSs and QSSs,
the distribution of $N_H$ should mirror the  spatial distribution
of SSSs along the line of sight within the galaxy. That is, since SSSs and QSSs
are generally expected to be evenly distributed along the line of sight,  
values of $N_H$ should range from approximately $N^G_H,$ the Galactic value
along that line of sight to a high of 
$\sim N^G_H + (z\, 3.08 \times 10^{18})/cos{\theta},$ where $z$ is the
scale height of the galaxy  
(in pc) 
and $\theta$ is the angle between the normal to the galactic plane and the 
line of sight. (We have assumed that, within the galaxy, there is on average one 
hydrogen atom per cubic cm.) For the galaxies we have studied,
this generally means that $N_H$ should range from a few times $10^{20}$ cm$^{-2}$
to a few times $10^{21}$ cm$^{-2}.$  Because absorption  
interferes with               
very soft source detection, we expect a bias toward values of $N_H$ on the
lower end of this range. Yet, the values of $N_H$ 
associated with the QSS fits summarized in Table 6 tend
toward
the high end.
This could imply that internal absorption is generally high in QSSs.
We note however that it is also possible that there is
a systematic bias in the fits, perhaps due to an
overestimate of the efficiency with which ACIS is detecting
soft X-ray photons. (See also \S 5.3.2.) 
Improvements in the low-energy calibration
are required to resolve this issue.

\subsubsection{Special Cases}

M101-102 is an ultraluminous SSS 
($L_X \sim 4\times 10^{39}$ erg s$^{-1}$).
No one-component model 
provides an acceptable fit. We therefore fit the spectrum with an absorbed
blackbody plus RS model. We found that the 91 eV blackbody component
contribute 85\% of the X-ray emission in 0.3-7 keV. 
This source has recently been studied by Mukai et al. (2002), 
Di\,Stefano \& Kong (2003a) and Kong \& Di\,Stefano (2004). 

In $2$ other cases, 
blackbody models did not
provide acceptable fits.
M101-53, a QSS, found by applying the ``$\sigma$'' condition, can be fit 
by a RS model with $k\, T_{RS}=0.64$ keV.
A composite of QSS-SNOH sources in M101 
requires a two-component model (RS plus power-law). The RS
temperature is 0.25 keV and the photon index of power-law is 2.4; the
thermal component contributes about 22\% of X-ray emission. 
We suspect that M101-53 and one or more of the SNOH sources
comprising the composite may be supernova remnants.

Four QSS candidates in M83 (M83-116, M83-117, M83-119 and M83-120) are
near the nucleus where the diffuse emission
contaminates the source spectra. Moreover, M83-116, M83-117 and M83-119
are near ($\sim 1''$) to each other while M83-120 has a nearby bright X-ray
source. These circumstances
make the spectral analyses difficult, and we therefore took special
steps to fit the  spectra of these sources. In each case, in addition to
the
source spectrum, we also extracted a spectrum from a source-free region in
an adjacent area with diffuse emission. We used this  latter as background. 
After background (diffuse
emission) subtraction, we fit the spectra with single-component models and
the results are listed in Table 6. Blackbody models with temperatures of
$\sim 140$ eV gave acceptable fits to M83-116 and M83-119. For
M83-117 and M83-120, the background appeared to be over-subtracted and no
single- or two-component model provided a good fit. We therefore considered
the source spectrum without background subtraction and fit a
two-component model (MEKAL + blackbody).
The MEKAL component represents the contribution of the diffuse emission
(see Soria \& Wu 2002) while the blackbody component corresponds
to the spectrum 
of the point source. Table 6 lists the best-fit results. The MEKAL plasma
temperature is roughly consistent with the value for the diffuse
emission found previously (Soria \& Wu 2002) while the blackbody
temperature is about 60 eV. The flux contribution of the blackbody
component is 82\% and 43\% for M83-117 and M83-120, respectively.

\subsubsection{Tests of the Selection Algorithm}

\noindent {\sl SSSs:\, }The spectral fits allow us to determine what sorts of sources are selected by our algorithm. As expected, the HR conditions are the most stringent. 
When applied to M101, M83, M51, and NGC 4697, they identified
sources that have $k\, T \laeq 100$ eV. Even for the ultraluminous
source in M101 (M101-102), which required a two-component fit,
fewer than $2\%$ of the energy was emitted in the form of
photons with energies above $1.5$ keV. In addition, the 
values of
$N_H$ among the HR-identified sources are 
distributed more-or-less uniformly between a few 
$\times 10^{20}$ cm$^{-2}$ and a few 
$\times 10^{21}$ cm$^{-2},$ roughly consistent with the expected distribution 
of obscuring gas. Thus, the HR conditions seem to identify sources
with the spectral characteristics expected of classical SSSs. 

The SSS category encompasses both SSS-HRs and SSS-$3\sigma$s. The SSS-$3\sigma$ category should consist of very soft sources with count rates too low to allow the straight HR conditions to be satisfied. In the 4 galaxies, no SSS-$3\sigma$ provided enough enough photons to allow a spectral fit. To carry out a fit we therefore considered a composite of 7 SSS-$3\sigma$ sources. Although this composite spectrum appears to be significantly harder than the SSS-HRs, both the formal and systematic uncertainties (the latter largely associated with creating the composite spectrum) are large. We therefore note that our results on SSSs are dominated by the bright SSS-HRs in our sample. 
\\

\noindent {\sl QSS:\, }
We were able to derive fits
for $10$ QSSs and $3$ QSScomposites identified by using other conditions.
As is to be expected, the photon statistics were generally
not quite as good for sources identified via the other
conditions, and the uncertainties in the spectral
parameters are greater.\footnote{There were a 
relatively large number of photons associated
with each of the sources near the center of M83, but the effects
of the diffuse emission introduce additional uncertainties.}  
The fits of two of these
spectra (M101-53, and the SNOH composite in M101) seem compatible with
what we might expect for SNRs. For the other spectra, 
the best-fit temperatures are 
generally higher than for sources in the HR category. 
Nevertheless, with just one exception, 
$k\, T < 175$ eV.

\noindent {\sl Comparison:\, }
The QSS spectra are generally harder and the values of $N_H$
appear to be systematically higher.   
In fact $7$ of the fits yield values of $N_H$ larger than the
largest value derived for HR-identified sources. In at least
$3$ cases, internal absorption would have to be very significant,
or the gas would have to be associated with a high-gas-density
region, in order for the derived values of $N_H$ to be physical.  
This could be a genuine selection effect, since some of these
categories were especially designed to allow us  to identify SSSs
located behind large gas columns, which could be associated with the system or which 
could be spread out along the line of sight. 
\\
 
\subsection{Luminosity}

\subsubsection{Luminosities}

When spectral fits are possible for only a small
fraction of sources in an external galaxy, the luminosity function is
usually computed by assuming that a single
factor can convert the count rate of
 each source into an estimated source luminosity.
The uncertainties can be estimated by using a set of 
spectral models (perhaps drawn from those that fit the 
high count-rate sources)
to compute independent conversion factors, and using them to derive
a range of physically plausible LFs.
For VSSs, however, the conversion from count rate to
luminosity depends critically on both the source
temperature and the value of $N_H.$ 
To gain insight into the distribution of luminosities we
can therefore rely only on the  sources for which we 
have spectral fits. In table 
$6$ we list the computed luminosities for each
source and composite for which we 
derived spectral fits. For each source we list both
a low value of the luminosity
(assuming that the galaxy's distance from us is the minimum
listed in Table 1) and a high value
(assuming that the galaxy's distance from us is the maximum
listed in Table 1). We have used these values to
compute $1$ LF for the associated VSSs using the lower distance estimate. For those sources that are
composites, we have assumed that the spectra of the sources contributing
to the composite are similar. We have therefore assigned to each source
a fraction of the total luminosity
that is equal to the fraction of the total counts contributed by that
source. In total, $40$ sources contribute to our
LF: $14$ single sources and $26$ sources whose luminosities
were determined by studying the spectrum of a composite
to which they contributed. 
The results are shown in Figure 2. 

While the function shown in Figure 2 displays some
interesting-looking features, it is not possible to compare
it directly to the LFs of galactic populations of harder
sources.
Perhaps the largest uncertainty is that we do not know how
well our assumed distribution of luminosities for
sources comprising composites reflects reality. In addition,
for some of the composites, the temperatures and luminosities
are not very well constrained. Finally, we do not
know how well this distribution reflects the characteristics
of the dimmer sources in our sample.  
Nevertheless there are a few noteworthy features. 
%
Among the sources in this sample, about $18\%$ have
$L < 10^{37}$ ergs s$^{-1}$ and 27\% of the sources have
$L > 10^{38}$ ergs s$^{-1}$ .  

Referring to Table 6, we find that 
the low-luminosity
sources ($L < 10^{37}$ ergs s$^{-1}$) tend to have
high temperatures, between $80$ and $170$ eV. 
The high temperature of the low-luminosity sources is a selection
effect--we simply would not have detected these sources had their 
temperatures been well under $100$ eV. It is, however, interesting that
there are low-luminosity sources with such high temperatures. One of
the predictions of nuclear-burning WD models is that
the luminosities and temperatures are larger for WDs of higher mass.
Low-L/high-T sources do not seem to fit this pattern.
Given the uncertainties in the spectral   parameters, 
and the assumptions made 
about the composite spectra, it may be too early to conclude that
the apparent low-L/high-T sources are not nuclear-burning white dwarfs (NBWDs). But the data
do suggest that these systems may be described by a different
physical model.

\subsubsection{Count Rates}

Figure 3 shows, for each galaxy,
the distribution of count rates for VSSs (red)
and for all other sources (black).
There are no obvious breaks in the distributions as we
approach lower count rates. This indicates that, if we had
longer and/or more sensitive observations, we would
detect both VSSs and other sources that are even dimmer
than the sources discovered so far.  

In general, there are fewer VSSs than non-VSSs in each galaxy.
This is especially true for NGC 4697 and M51. There is
nevertheless some preliminary evidence that count-rate distribution
 for VSSs in each galaxy is more sparsely populated at high-count-rates.
This makes the few high-count-rates SSSs particularly interesting.

\subsubsection{Ultraluminous SSSs}

One of the VSSs is clearly an ultraluminous X-ray source (ULX).
With a luminosity that appears to be larger than 
$4 \times 10^{39}$ ergs s$^{-1}$, M101-102 (a HR source) is clearly super-Eddington for a
neutron star. It cannot be explained by a NBWD model.
Three of the sources in NGC 4697 are also particularly luminous.
The $3$ sources comprising the HR composite in NGC 4697 are
roughly equally bright. The luminosities of the individual
sources would therefore be $\sim 10^{38}$ ergs s$^{-1}$ if
the distance to NGC 4697 is $11.7$ Mpc, but they could be
as high as $\sim 4 \times \sim 10^{38}$ ergs s$^{-1}$ if
the distance to NGC 4697 is $23.3$ Mpc. If the true 
luminosities are sub-Eddington for a $1.4 M_\odot$ object, then these systems
could include NBWDs that are both hot and bright because they
are approaching the Chandrasekhar limit.
Since super-Eddington luminosities can occur during short-lived states
(e.g., in Nova LMC 1990\#1; see Vanlandingham et al.\, 1999),
it will be important to study the time variability properties of such
luminous soft sources.

\subsection{Location}

Identifying very soft sources in external galaxies allows us to determine their locations relative to galaxy populations of young and old stars and also relative to the nucleus, spiral arms, and other galactic structures. This information can in turn, provide valuable clues to the natures of the sources.

\subsubsection{Near the Galaxy Centers}

It is impossible to detect the soft X-radiation from any SSSs/QSSs 
that may be near the center of the Milky Way.
Furthermore,
because the nature(s) of the sources are not well understood,
there were no predictions about whether we should find
SSSs/QSSs near galaxy centers.
In M51, we excluded the inner $15''$ because of possible
contamination from the central AGN. In each of the
other $3$ galaxies, we would have discovered any SSSs located
near the center. 
Only M83 has SSSs/QSSs within $100$ pc of the galaxy center.
The $3$ sources identified by the $\sigma$ conditions, 
M83-116, M83-117, and M83-120 are within
$45''$ of the nucleus. Because of possible contamination from the
diffuse emission, it will require further study to determine
if these are genuine point sources.     

All of the galaxies in our sample have VSSs within 1 kpc of the center.  The significance of the value of 1 kpc is that this is the distance a source moving at an average speed of 200 km s$^{-1}$ could travel in 5 million years. If a very soft sources is the stripped core of a tidally disrupted giant, then its maximum lifetime as a soft source should be approximately $10^3-10^6$ years (Di\,Stefano et al. 2001). 

\noindent {\sl M101:\ } 
There are $2$ SSSs and $2$ QSSs within 1 kpc; $2$ of these were selected
by the HR conditions, $1$ by HR1, and one by $3\sigma_1.$
There are  $5$ additional X-ray sources within $1$ kpc.

\noindent {\sl M83:\ }  
There are $9$ SSSs and QSSs within 1 kpc; $1$ HR,  $1$ $3\sigma$ source, $4$ $\sigma$ sources, and, $3$ HR$_1$. 
That same region contains a total of $24$ X-ray sources. 
We note however, that the {\it only} sources within $100$ pc of the
center are very soft sources, suggesting that it is important to
to verify or falsify
the point-like nature of these diffuse-emission-contaminated sources.
 
\noindent {\sl M51:\ } 
There is just one (QSS-$3\sigma_1$)
source within $1$ kpc of the center of M51, while there is a
total of $3$ X-ray sources in this region.

\noindent {\sl NGC 4697:\ } $1$ SSS and $4$ QSSs
($1$ HR, $3$ SNOH, and $1$ HR$_1$) are located within $1$ kpc of the
galaxy center. 
Only $14\%$ of all of NGC 4697's X-ray
sources are located within $1$ kpc  of the center.          

These results are summarized in Table 7, 
where we also compute a central concentration factor 
for both very soft sources
and all X-ray sources. Let $f_{SSS+QSS}$ 
($f_{all}$) represent the fraction of
very soft sources (all X-ray sources)
found within $1$ kpc of the galaxy's center.
In addition, let $f_{area}$ be the fraction of the galaxy's
total surface area comprised by a $1$ kpc disk.  The concentration
factor for SSSs, ${\cal F}_{SSS+QSS},$ is computed as follows. 

\begin{equation} 
{\cal F}_{SSS+QSS}={{f_{SSS+QSS}}\over{f_{area}}}    
\end{equation} 

Table $7$ demonstrates that X-ray sources are centrally concentrated
in all $4$ galaxies. It also demonstrates significant differences
among the galaxies, with relatively mild central concentration
(${\cal F}_{SSS+QSS}=15$) in M101, moderate concentration
(${\cal F}_{SSS+QSS}=50$) in M83 and M51, and heavy concentration
concentration
(${\cal F}_{SSS+QSS}=360$) in NGC 4697.  
Only in NGC 4697 are the SSSs significantly more centrally
concentrated than other X-ray sources. These results suggest that a more complete study is warranted. More galaxies need to be included and the fall off of optical light versus X-ray source density should be studied. Since the central regions are dense in sources of optical light, comparing the X-ray source distribution to that of optical light could provide insight into the mechanisms by which the X-ray sources are produced.
  
\subsubsection{In the Spiral Arms}

Superposing the positions of the SSSs and QSSs on optical images of each of
the $3$ spiral galaxies produces an impression that a large
fraction of the SSSs and QSSs are in the spiral arms. 
(See Figure 4  for all of the sources and Figure 5 for
those sources with spectral fits.) 
In most cases, populations
found in the   spiral arms are young populations, with ages on the
order of $10^8$ years. If, therefore, SSSs are primarily associated with
WDs orbiting low-mass ($< 2.5 M_{\odot}$) companions, we would not expect them to be a spiral arm population, but
to be spread across the face of the galaxies in a manner that
resembles the distributions of planetary nebulae (PNe) or recent novae. (Indeed, we expect some SSSs to be central stars of PNe or to be recent novae.)

Old populations or intermediate age populations (with current ages
greater than a few times $10^8$ years), should not be primarily
associated with spiral arms. Certainly we expect some of the
systems we have observed as SSSs to be old. Novae, for example,
should be scattered throughout the body of each galaxy. In addition,
because the donors in symbiotics are giants with masses typically
below $2\, M_\odot$ (Kenyon 1986), they should also be older than
$\sim 10^9$ years.
$4$ of $9$ Galactic SSSs with optical IDs are either novae
or symbiotics, as is $1$ of $3$ SMC
SSSs with optical IDs, and $1$ of $7$ LMC SSSs with optical IDs.
One SSS and $1$ QSS in M31 are tentatively identified with
symbiotics (Di Stefano et al. 2003), as is $1$ nova
(Nedialkov et al. 2002); these constitute the only optical IDs
of SSSs in M31.
Of the remaining $13$ SSSs with optical IDs in the Magellanic Clouds
and Galaxy, $9$ are candidates for the close-binary supersoft source
(CBSS) model. The donors in CBSSs are thought to main sequence 
or slightly evolved
stars with $M$ below approximately $2.5\, M_\odot.$
Whether the donor is slightly evolved or on the
main sequence, the ages of the systems at the time
mass transfer begins the SSS phase is typically larger than $10^8$ years. 
Thus, although detailed population synthesis compuations are needed
to quanitfy the relationship between location, level of activity,
and age, it seems clear that most WD SSSs (with the possible exceptions
of very hot central stars of planetary nebulae descended from
massive stars, or accretors of heavy winds from massive companions)
should not be part of young populations.

More convincing than a visual impression of an association
between the locations of VSSs and the spiral arms,  would be
a meaningful statistical correlation between the
positions of the VSSs and the positions of markers of young
populations, such as SNRs, OB associations, and HII regions.
Work is underway to study these  correlations in M101.
Below we mention some specific cases in which SSSs 
are close to optical markers in M101. The results  of a
statistical analysis will be reported elsewhere.
 
Nine of the VSSs (7 SSSs and 2 QSSs) are within $10''$ 
($\sim 240$ pc, at the distance to M101)
of
supernova remnants (SNR; Matonick \& Fesen 1997).  While at this point we cannot make a 
convincing
identification between the VSSs and the nearest SNR to each,
it is likely that a subset of these $9$ VSSs may in fact be
associated with SNRs, especially since all of the $7$ SSSs are within $2''$
($\sim 48$ pc) 
of an SNR. Astrometric study is underway to 
assess the likelihood of genuine identifications.
Because all $9$ sources have relatively
low count rates, we do not have spectra. 

In addition, $19$ distinct VSSs (11 SSSs and 8 QSSs) are within $10''$ of 
an HII region (Scowen et al. 1992),
with one SSS located near $3$ HII regions, and $2$ VSSs (1 SSS and 1 QSS) 
each
located near $2$ distinct HII regions.    
We have spectra for $4$ of these sources. M101-104 and M101-51,
each identified through the strong HR conditions,  
are among the softest sources in our sample, with $k\, T$ of $67$ eV
and $50$ eV, respectively. M101-114 was identified through the MNOH
condition, and is well fit by a $k\, T = 124$ eV model; the
uncertainties in the fit yield a maximum value of $k\, T$ of $\sim 150$ eV.
We have conjectured that M101-53, identified through the $\sigma$ condition,
may be an SNR, since the best-fit model was an RS model  with 
$k\, T_{RS} = 0.64$ keV.  Furthermore, $1$ QSS is within $10''$ of a PN 
(Feldmeier et al. 1997).   

If indeed we can establish a spatial correlation between VSSs 
and markers of young stellar populations, and if this
correlation is significantly larger than the correlations between 
PNe (and novae) and markers of young stellar populations,
then we will be forced to conclude that a substantial fraction
of VSSs are members of relatively young stellar populations. 

We must then consider whether the VSSs we are identifying
are dominated by SNRs. 
It should be possible to distinguish SNRs
from VSSs which are X-ray binaries by taking
multiple X-ray images; SNRs, whose emission emanates from
regions extending over several parsecs, should not be variable on 
time scales of
months to a few years, while most other VSSs are highly variable over
these time scales (Kong \& DiStefano 2003a; Di\,Stefano \& Kong 2003b; 
\rd\ et al. 2003; Fabbiano et al. 2003). 
We note that some SSSs which are X-ray binaries are
expected to be 
associated with ionization nebulae.
If the nebulae are not too dim or too diffuse,
VSSs of medium $T$ (likely to be SNRs) and VSSs
of lower-T (more similar to SSSs in the Milky Way and
Magellanic Clouds)
may be distinguished by various 
sets of line ratios (see, e.g.,
Rappaport et al. 1994, \rd , Paerels \& Rappaport 1995, 
Chiang \& Rappaport 1996).

Even before such follow-up observations are conducted,
we can conclude that it is unlikely that the VSS 
populations we have identified
are dominated by SNRs.   
First, this would be counter to the trend  discovered by 
Long et al. (1981) for the Magellanic Clouds.
 The $2$ SSS binaries, CAL 83 and CAL 87 had hardness ratio
[(H-S)/(H+S), where S includes photons between $0.15$ and $1.5$ keV
and H includes photons between $1.5$ and $4.5$ keV] of $-0.9$
and $-0.7,$ respectively. No SNR is as soft as CAL 83 and only $4$
are as soft or softer than CAL 87. X-ray luminosities are
quoted for $3$ of these SNRs, and they range from
$2 \times 10^{35}$ erg s$^{-1}$ to $4 \times 10^{35}$ erg s$^{-1}$.
They would not be detected in any of the galaxies we have
studied. Although it may be the case that
some SNRs have both spectra and luminosities more typical
of SSS X-ray binaries (see, e.g., Kong et al. 2002; Kong et al. 2003; Swartz
et al.\ 2002), 
the survey of the Magellanic
Clouds is consistent with a general trend in which
SNRs become larger, softer, and less luminous with time.  
The second reason we doubt that the VSS population we identify 
is dominated by SNRs, is that the total populations of VSSs in the 
Galaxy, the Magellanic Clouds, and in M31 are certainly not dominated by
SNRs. In M31 (Di\,Stefano et al. 2003), the VSSs identified by our
criteria are highly variable on time scales of months to years,
so that the X-ray emission cannot be coming from an extended region.
(See the SSS catalog [Greiner 2000] for a summary of the properties
of SSSs in the Galaxy and in the Magellanic Clouds.)

\section{The Size of Galactic SSS Populations}

The SSSs we have detected in each galaxy comprise a small fraction of all
of the VSSs within the galaxy. We detect primarily the hottest, brightest
VSSs. 
This effect becomes even more pronounced for distant galaxies
and/or for (a) galaxies located along lines of sight
with large values of $N_H,$ (b) galaxies with large internal gas components,
or (c)  spiral galaxies with large inclination angles. 
 
To estimate the size of the total population, we must determine
what fraction of all the sources we can (1) detect and (2) identify
as VSSs. The second requirement is more restrictive,
as typically more than $14$ photons are needed.
Assuming that at least $14$ photons are needed, 
we can compute, for each temperature
and each value of $N_H,$  
the minimum luminosity of a detectable VSS in each galaxy.
We have done these  computations for each value of $T$ and $N_H$
used in the PIMMS simulations described in paper 2 (\rd\ \& Kong 2003b).

If we knew the true distribution of VSS properties ($L$ and $T$,
assuming that thermal models apply) and the spatial distribution
of gas within
each galaxy, we would be able to 
compute the
true fraction of VSSs we should have detected and identified
in each galaxy. We could even check whether the properties
of the sources we actually do detect  match those of the predictions.
Unfortunately, however, we do not know enough about VSSs  to
make reliable first-principles calculations possible.

What we can say is that galactic VSS populations can 
productively be viewed as having $3$ components.

\subsection{The ``classical'' SSSs} 

The first component mirrors the SSSs studied by ROSAT to
first establish the class. 
A significant  fraction of these 
are candidates for binaries which include a NBWD. Distributions
of source properties for this class were computed by Rappaport, Di\,Stefano \& Smith (1994; RDS) and later used by \rd\ \& Rappaport (1994) to
estimate the fraction of SSSs that the {\sl ROSAT All-Sky Survey} 
had discovered in the Galaxy, M31, and the Magellanic Clouds. 
Although specifically constructed to mimic the distributions
of CBSS properties, the RDS distributions  seemed to
fairly well represent the properties of all SSSs discovered
during the early 1990s. (Note however, that 
the source properties were only established for about a dozen
sources and that they
were poorly constrained.)   

We have used the same distribution to compute the fraction  
of all sources like those discovered in the early 1990s
that we would have detected and identified in 
M101, M83, M51, and NGC 4697. To accomplish this 
we constructed a template of sources 
from the PIMMS simulations described
in \S 3 of \rd\ \& Kong (2003a). Each source was characterized by a temperature (with
$k\, T$ ranging from $25$ eV to $175$ eV), and a value of $N_H$.

We then carried out a simulation in which we
 seeded each galaxy with the RDS distribution.
The galaxy's distributions of $N_H$ were modeled in a very simple
way. 
For NGC 4697 we assumed a column of $4\times 10^{20}$ cm$^{-2}$
for each source. 
For M101 and M83 we assumed that an SSS has an equal probability
of lying behind a column of $4\times 10^{20}$ cm$^{-2}$ 
or of $1.2 \times 10^{21}$ cm$^{-2}$.
 For M51, which has a higher 
inclination angle, we assumed that three values of $N_H$
are equally likely: $4\times 10^{20}$ cm$^{-2}$, 
$1.2 \times 10^{21}$ cm$^{-2}$, and $6.2 \times 10^{21}$ cm$^{-2}$.  

For each galaxy, we considered each source in the RDS distribution.
Comparing its temperature to the temperature of each source
in the template (which is characterized by $T$ and $N_H$), 
we computed the luminosity the source would
need to have in order to yield $14$ counts. If the actual
source luminosity was this high or higher, we  
counted the source as detected. 

The results are summarized in Figure 6 and in the paragraphs below.
The top panel of Figure 6 shows one of the ``parent" distributions
of ``classical'' SSSs. The panels below show the members of that
parent distribution that would be detected in M101,
M83, M51, and NGC 4697, respectively. We have assumed that the distance
to each galaxy is the average of the highest and lowest values
given in Table 1.  In all of these external galaxies, 
we can detect only the hottest and brightest sources in
the distribution. This effect is, however, very pronounced
in M51 and especially in NGC 4697, where we can detect
only the high-L, high-T tip of the distribution.
 
\noindent {\sl M101 and M83:\ } The fraction of sources we would 
detect ranged from $2\%$ to $5\%$, with a median value of $3\%$.
This implies that each ``classical'' SSS detected represents
between $20$ and $50$ SSSs, with a median expectation value
of $\sim 33$. 

\noindent {\sl M51:\ } The fraction of sources we would
detect ranged from $0.1\%$ to $0.25\%$, with a median value of $0.17\%$.
This implies that each ``classical'' SSS detected represents
between $400$ and $1000$ M51 SSSs, with a median expectation value
of $\sim 590$. 
 
\noindent {\sl NGC 4697:\ } The fraction of sources we would
detect ranged from $0.05\%$ to $0.2\%$, with a median value of $0.09\%$.
This implies that each ``classical" SSS detected represents
between $500$ and $2000$ NGC4697 SSSs, with a median expectation value
of $\sim 1100$.

For each galaxy we now have to estimate the number of classical
SSSs detected. Given the limited information we have, it is not presently
possible to do this in a reliable way. We could, for example
attempt to estimate the number of
SSSs that seem different from the ``classical'' profile and
subtract these from the total number of VSSs in each galaxy, to arrive 
at an upper limit. We do not, however, have enough information
about individual sources to make this assessment.
On the other hand, the HR category does seem to select
sources with temperatures in the appropriate range. 
Some of these, however, have luminosities that are too large. 
To make a conservative estimate, we utilize {\it only} sources in
the HR category, using the total number, $N_{HR},$ of such sources to
derive a conservative upper limit, and half of the total
in HR to derive a lower limit. This is conservative because it
excludes the majority of SSSs in each galaxy. It assumes 
that the sources identified outside of the HR
condition are most likely to represent types of SSSs
that are not common in the solar neighborhood
 or in the  Magellanic Clouds. 

We multiply the median expectation value for each galaxy by
$N_{HR}/2$ and by $N_{HR}+N_{3\sigma}$ to derive a range
for the number of ``classical'' SSSs. For M101 the range is
from $330$ to $660$. For M83 the range is
from $250$ to $500$. For M51 the range is
from $1500$ to $3000$. For NGC 4697 the range is from
$2200$ to $4400$.        

There are $2$ important caveats. First, for
the galaxies that are farthest from us, these estimates
would be overestimates, if
none of the SSSs we detect actually have the classical profile
but are instead more luminous and hotter. This 
would not imply that NGC 4697
doesn't have a large ``classical'' SSS population, rather that we cannot
estimate the size of that population based on the data.
The second caveat is that, for the spirals that may be closer
to us, M101 and M83, the conservative estimates may be underestimates.

\subsection{Nova-like variables and other high $\dot m$ CVs} 
The second class of SSSs we expect to inhabit distant galaxies
are the low-L ($10^{35}-10^{37}$ erg s$^{-1}$, low-T ($k\, T < 40$ eV)
sources that may be associated with nova-like
variables or CVs with particularly high accretion rates--
 systems like V751 Cyg (Greiner et al. 1999; Patterson et al. 2001), 
and V Sge (Greiner  \& van Teeseling  1998).
  These systems would not be detected by the observations carried
out so far, so the observations we have used here cannot
constrain the sizes of these populations in the $4$ galaxies we
study.
Nevertheless, 
some reasonable assumptions can be made. 
First, it is very likely that each of these galaxies houses
a population of CVs. 
If the old stellar populations of the $4$ galaxies in our
study have similar characteristics to those of the Galaxy,
we might expect there to be $\sim 10$ 
times as many low-L, low-T SSSs ``on''
at any given time, 
as there are SSSs with $L> 10^{37}$ erg s$^{-1}$.

\subsection{New Classes of Very Soft Sources}

It is clear that, among the VSSs we have identified in 
M101, M83, M51, and NGC 4697, there are sources unlike any
found to date in the Galaxy, Magellanic Clouds, or in M31.
At this point, it is difficult to estimate the size of these ``new'' 
populations. 
The QSSs comprise one new class. Many are hotter 
than the canonical SSSs,
and may therefore be easier to detect; the catch, however,
is that many 
are near the detection limits. There could be many more
below the detection limits.
Among the SSSs, we also find low-$L$ sources, some with luminosities for their estimated temperatures
so low that they may not be consistent with WD models.
Most likely, the systems detected so far
simply represent the most detectable members of the ``new'' SSS 
populations. 

To be able to estimate the size of these new populations we need 
physical models for them, so that population synthesis simulations
can produce distributions with which to seed galaxies.
We also need further observations of these and of more nearby galaxies,
to explore lower luminosities and temperatures.

\section{Discussion}

The work described in this paper has served three purposes. First, 
it tests the selection algorithm presented in the companion paper.
By fitting spectra to those sources providing
adequate numbers of counts, we find that the criteria
uniformly
select soft sources. As expected from the tests conducted on simulated data (Di\,Stefano \& Kong 2003a), those VSSs fit by blackbody spectra had $kT < 250$ eV, while all VSSs emitted less than 10\% of their flux in photons with energies greater than 1.5 keV. Second, we establish the existence of QSSs. Sources with kT on the order hundreds of eV and $L > 10^{36}$ erg s$^{-1}$ appear not to have been observed before. By now, observations in several nearby galaxies, including M31 (\rd\ et al. 2003a) and M104 (\rd\ et al. 2003b) verify the result. Finally, investigations 
have also yielded interesting information about extragalactic SSSs,
providing some first steps toward answering the $7$ questions
posed in Di\,Stefano \& Kong (2003a). We summarize this progress below.

(1)  {\sl What are typical galactic populations of very soft sources?} 
It is too soon to claim an answer to this question.
We can say, however, the SSSs and QSSs seem to exist in virtually
every galaxy. (Ongoing analyses of additional galaxies lend
further support to this conclusion.)
We have so far carried out only a crude population analysis
of the $4$ galaxies studied here. In each case though, the 
data is consistent with a population of at least $\sim 500$
VSSs with $L > 10^{37}$ erg s$^{-1}$ active at the 
time of the observation. The true underlying population of
such luminous sources could be several times larger. 
Observations so far provide evidence for galactic populations
of low-luminosity VSSs ($L < 10^{37}$ erg s$^{-1}$), but do not 
constrain its size.
With regard to the potential of
VSSs to ionize galactic ISMs, we note that even the
most conservative estimate of VSS population for these
galaxies yields more than $10^{40}$ erg s$^{-1}$
of highly ionizing radiation released by SSSs in typical galaxies.
Although the luminosity is smaller than that associated with
young stars, photons from VSSs may dominate at high energies. 

(2)  
{\sl Are any spiral galaxy parameters
related to the relative sizes of SSS and QSS populations?} 
Among the spiral galaxies in this study, there is not enough variation
among the values of $M_B$ to allow a search for correlations
between $M_B$ and the true size of the SSS/QSS population. 
The ratio of the number of detected SSSs and QSSs to the total number of 
detected X-ray sources does vary systematically with inclination
($39\%$, $37\%$, and $30\%$ for M101, M83, and M51, respectively).
If we limit consideration to sources selected by the HR conditions,
which seem to be the softest sources selected,
the trend appears even stronger
($17\%$, $11\%$, and $7\%$ for M101, M83, and M51, respectively). 
This  might suggest similar sized populations in each of
these galaxies. 

(3) 
{\sl Do elliptical galaxies house large SSS/QSS populations?} 
The existence of SSSs and QSSs in NGC 4697 demonstrates that there
are SSSs in ellipticals. 
(See also Sarazin, Irwin, \& Bregman 2001.)
It is difficult to estimate the size of the 
underlying SSS/QSS populations in galaxies like NGC 4697,
largely because we do not yet have good models for the 
characteristics of the SSS/QSS population in these galaxies. 
We have demonstrated, however, that, if the sources we detect
in NGC 4697 are similar to the SSSs in the Milky Way and Magellanic
Clouds that were used to define the class, its population of
SSSs is large--about 1250 ``classical'' SSSs per detected
example of a ``classical'' SSS in NGC 4697. It is possible,
however,  that many of the $14$ detected VSSs are
examples of new types of sources.  Finally, we note that
the data so far are consistent
with the hypothesis that a significant portion of the
diffuse soft emission  found in elliptical galaxies may be
due to unresolved SSSs (Fabbiano, Kim, \& Trinchieri 1994).

(4) {\sl Within spiral galaxies, what are the relative populations
of SSSs/QSSs in the galaxy bulges and disks?}  
In the $3$ spirals we have studied, $f_{SSS+QSS}$,
the fraction of SSSs and QSSs
located within $1$ kpc of the galaxy center is the same
as $f_{all}$, the fraction of all X-ray sources 
located within $1$ kpc of the galaxy center. The values of
$f_{SSS+QSS}$ vary significantly among the spiral galaxies, however.

(5) 
{\sl Do galaxies with massive central black holes
have more SSSs/QSSs located within 1 kpc of the nucleus than
comparable galaxies without massive central black holes?} 
To answer this question, we have to rely on the dynamical
measurements of the central supermassive BH. 
M51 is a Seyfert 2 galaxy with a low-luminosity AGN (see, e.g.,
Ho, Filippenko, \& Sargent 1997) and it is very likely that there is
a supermassive,  
$\sim 10^7 M_{\odot}$ (Hagiwara et al. 2001) BH in the
center.  M83 might also harbor a $\sim 10^7 M_{\odot}$
BH (Thatte et al. 2000). 
The concentration factors for SSSs/QSSs in both 
M83 and M51 are high, approximately $50$ for each galaxy.
The concentration factors for all sources  
 in each galaxy is comparable, but somewhat smaller. 
\\
\noindent M101 might have a $< 10^6 M_{\odot}$
supermassive BH (Moody et al. 1995). The relative values
of the concentration factors among the spiral galaxies
suggests that
the concentration of SSSs/QSSs, 
and in fact the concentration of all X-ray sources, 
scales with the mass of the
central supermassive BH. It is interesting, but not at 
this point statistically significant, that the concentration
factors are somewhat higher for SSSs/QSSs.  
\\
\noindent 
NGC 4697 appears to have the highest-mass central BH of any of the 
galaxies in our sample ($1.2 \times 10^8 M_\odot$).
It is also the galaxy that has the highest concentration factor for
all X-ray sources (${\cal F}_{all} = 140),$ 
the highest concentration factor for SSSs
(${\cal F}_{SSS+QSS} = 360$), and the largest difference 
between the central concentration of SSSs and other X-ray
sources. From this  it seems that interactions near the
galaxy center may be playing an important role in creating X-ray sources,
and that this is especially true for SSSs/QSSs. (An alternative
hypothesis is that the characteristics of the stellar
population near the center differ from those elsewhere in the 
galaxy.)   
A variety of interactions could enhance the formation of
X-ray binaries near the center of the galaxy; it may
be that the additional enhancement for SSSs is due to
the presence of the cores of stars that have been tidally
disrupted by the central BH.

(6) 
{\sl For all galaxies, are the positions of SSSs and QSSs
correlated to the positions of other objects,
such as HII regions, planetary nebulae, supernova
remnants, or globular clusters?}   
In the $3$ spiral galaxies we have studied, the SSSs seem 
to be concentrated in the spiral arms.
Ongoing work on SSSs and QSSs in M101 indicates that they
may be preferentially located near 
HII regions and SNRs, suggesting that some SSSs/QSSs in M101 may be related to
young populations.

(7) Are SSSs significant contributors to the rates of
Type Ia SNe?
It is too soon to say. To use the data to answer this question,
we need to estimate the fraction of the SSSs in each galaxy
that are accreting WDs. This is difficult to do, even in the Magellanic
Clouds. There are, however, some signatures which can help.
First, the WD models developed to date invoke donor stars that 
too old to be concentrated in the spiral arms; the fraction
of sources located away from the spiral arms and from regions with recent star formation can provide clues about the size of the NBWD population.
  Second,
some of the accreting WDs that may be
Type Ia progenitors should be surrounded  by large, distinctive 
ionization 
nebulae (\rd\ 1996).\footnote{The reasons are the following. (1) SSSs
emit enough photons to ionize large ($> 10$ pc) regions of the 
surrounding ISM, if the number density in their surroundings is
moderate, on the order of $1$ atom cm$^{-3}$  (Rappaport, Kallman,
\& Malina 1994). (2) These nebulae have distinctive
optical emission signatures that can be significantly
different from other nebulae, such as PNe and SNRs 
(Rappaport \& Chiang 1996; \rd\, Paerels, \& Rappaport 1996). 
(3) These nebulae could be
detectable for much of the lifetime of the SSS state
($\sim 10^7$ yrs), as long as the duty cycle is $> \sim 1\%$
(Chiang \& Rappaport 1996).} In nearby galaxies where optical
counterparts can be identified, we may be able to detect some 
of the distinctive disk features of close binary SSSs ($P_{orb}< 3$ days) or 
to detect the donor stars in wide binary SSSs ($P_{orb}$ generally is more than 100 days).  

\subsection {New Questions}

\noindent (8) {\sl What are the ultraluminous SSSs?} 
A natural suggestion is that they are BHs. 
In fact, if there are BHs with masses filling the gap
between the $\sim 10 M_\odot$ BHs thought to be stellar
remnants, and the $> 10^6 M_\odot$ BHs found   at the
centers of galaxies, then SSS behavior is predicted for a wide range
of masses, source luminosities, and efficiency factors.
If BHs with masses in this intermediate range
do not exist, or are not found in binaries, then the
explanation of ultraluminous  SSSs may be more complicated.  

\noindent (9) {\sl What are the 
low-luminosity high-temperature VSSs?}  
The data provide a hint of a population of sources with
temperatures near or over $100$ eV, but with luminosities
between $10^{36}$ erg s$^{-1}$ and $10^{37}$ erg s$^{-1}$.
NBWDs at this temperature are generally expected to
be more luminous. If evidence for this high-T/low-L
SSS population is found in other investigations, 
a new model for some SSSs will be required. At present, there is no
``canonical'' model for QSSs.   
Intermediate-mass BHs provide one possible explanation.  Consider, e.g., a 200 eV source with $L_X\sim 5\times10^{37}$ erg s$^{-1}$. This could correspond to a BH of roughly 50 M$_\odot$ accreting at a rate of roughly 1\% $\dot{m}_{Edd}$. (Note that if the bolometric luminosities were significantly smaller, the source might not be expected to be in the soft state.)

\noindent (10) {\sl Is there a significant population of quasisoft 
sources and, if so, what is the nature of the sources?}
Some of the SSSs and QSSs may
be SNRs. If so, we expect the flux we measure in observations 
taken during the next few years to be comparable to the values
already measured; some may be associated in optical nebulae.
It will be important to search for
 quasisoft sources that are not SNRs. 

\noindent (11) {\sl If a subset of VSSs are young ( $< 10^8$ yrs),
what are they?}
Some of them may be SNRs, but we have argued (\S 5.4.2)
that the majority of them are not SNRs. Some may be BHs,
some may be neutron stars, perhaps analogs of RX J0059.2-7138.
It is even possible that some may be NBWDs. One scenario we
propose (R. Di\,Stefano, in preparation) is one in which a primordial 
binary with $2$ 
relatively massive stars ($> 3\, M_\odot$) evolves into a 
binary with a massive WD in orbit with a massive star.
Even if the primary never fills its Roche lobe, the system
could become a symbiotic which experiences an SSS phase.
If the primary {\it will} fill  
its Roche lobe, an epoch of irradiation-driven winds prior to contact
could also lead to SSS behavior. Finally, in some of these
wind-driven systems, the primary mass may be decreased
enough that mass transfer is not dynamically unstable when
contact is established, and the SSS behavior can continue.
  
\subsection{Prospects}

{\it Chandra} observations are beginning to help us answer
key questions about SSSs and to demonstrate that there
are new horizons in SSS research. An unanticipated result of this research is the discovery that galaxies contain a significant number of X-ray sources that appear to be hotter than the canonical SSSs that established the class. Discovering the nature of these QSSs presents a fresh challenge.
Continuing observations with both {\it Chandra} and {\sl XMM}
should lead to important results.
With the publication of this paper we will make our code
available to other researchers, and hope that it can
facilitate comparative studies of the VSS populations of different
galaxies.

\begin{acknowledgements}

We are grateful to Pauline Barmby, Mike Garcia,
Jochen Greiner, Roy Kilgard, Miriam Krauss, Jeff McClintock, Koji Mukai,  
Paul Plucinsky, Will Pence, Andrea Prestwich, Frank Primini, Roberto Soria, Douglas Swartz, Harvey Tananbaum, Ben Williams, Kinwah Wu,
and Andreas Zezas 
for stimulating discussions and comments.   
We are grateful to Paul Green and the ChaMP collaboration
for providing
blank field source data. 
This research has made use of the
electronic catalog of supersoft X-ray 
sources available at\\
 URL http://www.mpe.mpg.de/$\sim$jcg/sss/ssscat.html
and maintained by J. Greiner.    
RD would like to thank the Aspen
Center for Physics for providing a stimulating environment,
and the participants of the 2002 workshop on  
{\it Compact Object Populations in External Galaxies}  
 for insightful comments.  AKHK acknowledges support from the Croucher Foundation. 
This work was supported by NASA under AR1-2005B, GO1-2022X,
and an LTSA grant, NAG5-10705.  
\end{acknowledgements}

\clearpage

\begin{deluxetable}{lcccccccccc}
\tabletypesize{\tiny}
\tablewidth{0pt}
\tablecaption{Summary of sample galaxies}

\tablehead{
Name& Type& Distance & $D_{25}$ & $M_B$ & Inclination & $N_H$ & {\it
Chandra} & $\log(M_{HI})$ & $M_{BH}$ &$N_{VSS}/N_{total}$ 
\tablenotemark{a} \\
    &     &  (Mpc)   &  (arcmin)&       &             &
($10^{20}$cm$^{-2}$) & exposure (ks)&$(M_{\odot})$ &$(\times 10^{8}
M_{\odot})$ & }
\startdata

M101 & Sc & 5.4, 6.7$^1$ & 23.8 & -20.45 & $0^{\circ}$ & 1.2 & 94.4&
10.05 &$< 0.01$ $^2$& 53/118\\
M83 &  Sc & 4.7, 4.57$^3$ & 11.5 & -20.31 & $24^{\circ}$ & 3.8 &
49.5&10.01 & 0.1\,$^4$ & 54/127\\
M51  & Sc\tablenotemark{b} & 7.7, 8.4$^5$ & 13.6 & -20.75 & $64^{\circ}$ & 1.6 &
14.9& 9.46 & 0.1\,$^6$ & 23/72\\
NGC 4697 & E & 23.3, 15.9$^7$, 11.7$^8$ & 7.1 & -21.67 & $44^{\circ}$\tablenotemark{c} &
2.1 & 39.3 & 0.53 & 1.2$^9$ & 19/91\\
\enddata
\tablecomments{All data are from Nearby Galaxies Catalogue (Tully 1988)
unless specified.}

\tablenotetext{a}{Ratio of number of VSSs to total number of X-ray
sources.} 
\tablenotetext{b}{Seyfert 2 galaxy.}
\tablenotetext{c}{For elliptical galaxies
the inclination is given by $3^{\circ}$ +
acos\,$\left(\sqrt{((d/D)^2 - 0.2^2)/(1 - 0.2^2)}\,\right)$, where d/D
is the axial ratio of minor to major diameter (Tully
1988). 
This inclination angle is generally unrelated to the value of $N_H$.} 

\tablerefs{1: Freedman et al. 2001; 2: Moody et al. 1995; 3: Karachentsev et al. 2002;
4: Thatte et al. 2000; 5: Feldmeier
et al. 1997; 6: Hagiwara et al. 2001;  7: Faber et al. 1989;
8: Tonry et al. 2001; 9: Ho 2002}

\end{deluxetable}

\begin{deluxetable}{lccccccccccc}
\tabletypesize{\scriptsize}
\tablewidth{0pt}
\tablecaption{SSS and QSS source list of M101}

\tablehead{
\multicolumn{1}{c}{Object}& R.A.& Dec. & \multicolumn{2}{c}{Soft} &
\multicolumn{2}{c}{Medium} & \multicolumn{2}{c}{Hard} & HR1\tablenotemark{a} & HR2\tablenotemark{b} & Note\tablenotemark{c}\\ 
\cline{4-5} \cline{6-7} \cline{8-9}\\
 & (h:m:s)& $(^{\circ}:\arcmin:\arcsec)$ &\colhead{Counts} & \colhead{S/N}
& \colhead{Counts} &
\colhead{S/N} &\colhead{Counts} & \colhead{S/N} & & &}
\startdata

SSS-HR \\
M101-15(10) & 14:02:59.4&+54:20:42&   31.3 &   5.3 &   1.3 &   0.6 &   0.0 &   0.0 &  -0.9(-0.8)  & -1.0(-0.9)\\
M101-18(13*) & 14:03:01.1&+54:23:41&  169.6 &  12.9 &   0.0 &   0.0 &   0.1 &   0.1 &  -1.0(-1.0)  & -1.0(-1.0)\\
M101-21(16*) & 14:03:02.5&+54:24:16&   74.2 &   8.3 &   0.0 &   0.0 &   0.0 &   0.0 &  -1.0(-0.9)  & -1.0(-1.0)\\
M101-34(27) & 14:03:07.9&+54:21:23&   35.7 &   5.8 &   1.8 &   0.8 &   0.0 &   0.0 &  -0.9(-0.8)  & -1.0(-0.9)\\
M101-37(30*) & 14:03:08.5&+54:20:57&    29.3 &   5.0 &   1.0 &   0.4 &   0.4 &   0.0 &  -0.9(-0.8)  & -1.0(-0.9)\\
M101-38(31) & 14:03:08.6&+54:23:36&   20.7 &   4.1 &   0.0 &   0.0 &   0.0 &   0.0 &  -1.0(-0.9)  & -1.0(-1.0)\\
M101-43(36) & 14:03:10.6&+54:21:26&   18.2 &   3.8 &   0.0 &   0.0 &   0.0 &   0.0 &  -1.0(-1.0)  & -1.0(-0.9)\\
M101-50(43) & 14:03:13.2&+54:21:57&   31.3 &   5.3 &   0.0 &   0.0 &   0.0 &   0.1 &  -1.0(-1.0)  & -1.0(-0.8) & SNR\\
M101-51(45*) & 14:03:13.6&+54:20:09&  219.1 &  14.6 &   0.3 &   0.1 &   0.0 &   0.0 &  -1.0(-1.0)  & -1.0(-1.0) & HII\\
M101-55(48) & 14:03:13.9&+54:18:11&   34.3 &   5.6 &   1.4 &   0.6 &   0.0 &   0.0 &  -0.9(-0.8)  & -1.0(-0.9)\\
M101-78(73) & 14:03:22.6&+54:20:38&   34.3 &   5.4 &   1.5 &   0.6 &   1.0 &   0.4 &  -0.9(-0.8)  & -1.0(-0.8) & HII\\
M101-80(75) & 14:03:24.0&+54:23:37&   68.7 &   8.0 &   0.8 &   0.3 &   0.0 &   0.0 &  -1.0(-0.9)  & -1.0(-0.9) & HII\\
M101-97(92) & 14:03:29.8&+54:20:58&   71.9 &   7.9 &   0.0 &   0.0 &   0.0 &   0.0 &  -1.0(-0.9)  & -1.0(-0.9)\\
M101-101(96*) & 14:03:31.9&+54:23:23&   30.9 &   5.1 &   0.0 &   0.0 &   0.0 &   0.0 &  -1.0(-0.9)  & -1.0(-0.9)\\
M101-102(98) & 14:03:32.3&+54:21:03& 8512.0 &  92.2 & 658.4 &  25.6 &  68.0 &   8.0 &  -0.9(-0.9)  & -1.0(-1.0)\\
M101-104(99*) & 14:03:33.3&+54:18:00&  227.8 &  14.6 &   0.3 &   0.1 &   1.1 &   0.3 &  -1.0(-1.0)  & -1.0(-1.0) & HII\\
&&\\
SSS-$3\sigma$ \\
M101-5 & 14:02:51.7&+54:19:33&   13.0 &   3.2 &   1.7 &   0.8 &   0.0 &   0.0 &  -0.8(-0.6)  & -1.0(-0.7) & SNR\\
M101-11(8*) & 14:02:57.0&+54:19:48&   34.2 &   5.6 &   0.0 &   0.0 &   2.5 &   1.0 &  -1.0(-0.9)  & -0.9(-0.8)\\
M101-16(11) & 14:03:00.5&+54:20:02&   31.9 &   5.2 &   0.0 &   0.0 &   4.2 &   1.5 &  -1.0(-0.9)  & -0.8(-0.7) & SNR, HII\\
M101-17(12) &14:03:00.8 &+54:20:17 & 17.1 & 3.5 & 0.0 & 0.0 & 2.8 & 1.1 &-1.0(-0.8)&-0.7(-0.6)\\
M101-24 & 14:03:04.6&+54:20:37&   15.3 &   3.5 &   0.0 &   0.0 &   0.0 &   0.0 &  -1.0(-0.8)  & -1.0(-0.8)\\
M101-27(20) &  14:03:05.2&+54:23:11&   19.0 &   4.0 &   1.7 &   0.7 &   0.8 &   0.3 &  -0.8(-0.7)  & -0.9(-0.8)\\
M101-49(42) & 14:03:12.8&+54:19:01&   18.6 &   4.0 &   0.5 &   0.2 &   2.6 &   1.0 &  -1.0(-0.8)  & -0.8(-0.6) & SNR, HII\\
M101-57(52) & 14:03:14.6&+54:21:52&   17.9 &   3.8 &   1.9 &   0.8 &   0.0 &   0.0 &  -0.8(-0.7)  & -1.0(-0.8) & SNR\\
M101-60(55) & 14:03:15.0&+54:20:18&   18.0 &   3.7 &   0.0 &   0.0 &   0.5 &   0.2 &  -1.0(-0.8)  & -1.0(-0.8)\\
M101-72(68) & 14:03:21.1&+54:19:08&   21.2 &   4.2 &   1.7 &   0.7 &   0.0 &   0.0 &  -0.9(-0.7)  & -1.0(-0.8)\\
M101-76(72*) & 14:03:21.7&+54:20:24&   19.4 &   3.7 &   0.6 &   0.2 &   0.0 &   0.0 &  -0.9(-0.8)  & -1.0(-0.8) & HII\\
M101-85(80*) & 14:03:25.3&+54:21:13&   27.3 &   4.7 &   0.0 &   0.0 &   3.6 &   1.3 &  -1.0(-0.9)  & -0.8(-0.7) & HII\\
M101-86(81) & 14:03:25.8&+54:21:25&   18.6 &   3.6 &   0.7 &   0.3 &   0.0 &   0.0 &  -1.0(-0.8)  & -1.0(-1.0) & SNR, HII\\
M101-88(84) & 14:03:26.6&+54:20:43&   45.8 &   6.1 &   4.3 &   1.7 &   5.3 &   1.9 &  -0.8(-0.8)  & -0.8(-0.7) & SNR, HII\\
M101-92 &14:03:27.9 &+54:18:45 & 27.1& 4.2 & 3.5 & 1.3 & 3.5 & 1.1 & -0.8 (-0.7) & -0.8 (-0.7) & HII\\
M101-95(90) & 14:03:28.8&+54:20:59&   24.6 &   4.4 &   3.1 &   1.3 &   0.3 &   0.1 &  -0.8(-0.7)  & -1.0(-0.8)\\
&&\\
QSS-NOH\\
M101-77 & 14:03:21.7&+54:22:32&    9.4 &   2.5 &   0.7 &   0.3 &   0.5 &   0.2 &  -0.9(-0.6)  & -0.9(-0.6) & PN\\
M101-93(88) &14:03:28.6&+54:22:02&    12.8 &   3.0 &   1.1 &   0.4 &   1.3 &   0.5 &  -0.9(-0.7)  & -0.8(-0.6)\\
M101-106(101*) & 14:03:33.9&+54:20:11&   23.1 &   3.7 &   3.9 &   1.4 &   0.0 &   0.0 &  -0.7(-0.6)  & -1.0(-0.8) & SNR, HII\\
&&\\
QSS-SNOH\\
M101-2(1) & 14:02:49.1&+54:18:42&   15.4 &   3.1 &   8.7 &   2.7 &   0.0 &   0.0 &  -0.3(-0.3)  & -1.0(-0.8) & HII\\
M101-25(18) & 14:03:04.7&+54:19:25&   16.0 &   3.3 &   4.3 &   1.7 &   1.4 &   0.5 &  -0.6(-0.5)  & -0.8(-0.7) & HII\\
M101-61(56) & 14:03:15.5&+54:17:04&   53.7 &   6.4 &   9.8 &   2.7 &   0.0 &   0.0 &  -0.7(-0.7)  & -1.0(-0.9) & HII\\
\\
&&\\
QSS-HR$_1$\\
M101-3(2) &14:02:49.6&+54:19:59& 19.4 &   4.1 &   6.4 &   2.3 &   1.8 &   0.7 &  -0.5(-0.5)  & -0.8(-0.7)\\
M101-96(91) & 14:03:28.9&+54:21:49& 13.3 &   3.1 &   4.7 &   1.8 &   1.9 &   0.7 &  -0.5(-0.4)  & -0.8(-0.6) & HII\\

&&\\
QSS-$3\sigma_1$\\
M101-6(4) & 14:02:52.1& +54:19:47 & 12.5 & 2.9 & 0.4 & 0.2 & 2.3 & 0.9 & -0.9 (-0.7) & -0.7 (-0.6)\\
M101-12 &14:02:57.1 &+54:22:33 & 9.6 & 2.5 & 0.0 & 0.0 2.0 & 0.8 & -1.0 (-0.7) & -0.7 (-0.5)\\
M101-48(41) &14:03:12.7 &+54:17:44 & 11.6 & 2.7 & 7.1 & 2.4 & 2.4 & 0.9 & -0.2 (-0.2) & -0.7 (-0.5) & SNR\\
M101-58(53) &14:03:14.8&+54:21:10& 13.6 &   2.9 &   5.2 &  1.9 &   2.0 &   0.9 &  -0.5(-0.4)  & -0.8(-0.6)\\
M101-59(54) &14:03:14.8&+54:21:23&  11.2 & 2.9 &   4.9 &   2.0 &   1.7 &   0.7 &  -0.4(-0.3)  & -0.7(-0.6)\\
M101-87(82) &14:03:26.2 &+54:19:51 & 11.6 & 2.9 & 0.5 & 0.2 & 1.5 & 0.6 & -0.9 (-0.7) & -0.8 (-0.6) & HII\\

&&\\
QSS-$\sigma$\\

M101-53(45*) &  14:03:13.6&+54:19:09& 76.3 &   8.4 &  36.3 &   5.9 &   3.9 &   1.4 &  -0.4(-0.3)  & -0.9(-0.9) & HII\\
&&\\
QSS-FNOH\\
M101-1 & 14:02:48.8&+54:20:34&   11.9 &   3.0 &   3.0 &   1.3 &   0.0 &   0.0 &  -0.6(-0.5)  & -1.0(-0.8)\\
M101-8(6) &14:02:53.2 &+54:18:55 & 5.7 & 1.7 & 9.9 & 3.0 & 1.2 & 0.5 & 0.3 (0.2) & -0.7 (-0.4)\\
M101-19(14) &14:03:01.3 &+54:21:33 & 10.2 & 2.8 & 5.7 & 2.2 & 0.9 & 0.4 & -0.3 (-0.3) & -0.8 (-0.6)\\
M101-112 & 14:03:39.0&+54:16:56&  17.3 &   0.6 &  19.6 &   3.7 &   0.0 &   0.0 &  0.1(-0.3)  & -1.0(-0.7)\\
&&\\
QSS-MNOH\\
M101-114(107) & 14:03:41.3&+54:19:04&  280.1 &  14.5 &  64.4 &   7.2 &   1.4 &   0.3 &  -0.6(-0.6)  & -1.0(-1.0) & HII\\
M101-116(109) & 14:03:51.9&+54:21:49&   99.8 &   8.7 &  15.3 &   3.2 &   0.0 &   0.0 &  -0.7(-0.7)  & -1.0(-0.9)\\

\enddata

\tablecomments{The object ID in parentheses is from Pence et al. (2001);
SSSs selected by Pence et al. (2001) are noted by ``*''.}
\tablenotetext{a}{(M-S)/(M+S); value in parentheses is the maximum of HR1.}
\tablenotetext{b}{(H-S)/(H+S); value in parentheses is the maximum of HR2.}
\tablenotetext{c}{Entries indicate that an object of the type listed is within
$10''$ of a VSS; SNR: SNR catalog (94 objects in total) from Matonick \& Fesen (1997); HII: HII region catalog (248 objects in total) Scowen et al. (1992); PN: PN catalog (65 objects in total) from Feldmeier et al. (1997)}

\end{deluxetable}

\begin{deluxetable}{lcccccccccc}
\tabletypesize{\scriptsize}
\tablewidth{0pt}
\tablecaption{SSS and QSS source list of M83}

\tablehead{
\multicolumn{1}{c}{Object}& R.A.& Dec. & \multicolumn{2}{c}{Soft} &
\multicolumn{2}{c}{Medium} & \multicolumn{2}{c}{Hard} & HR1\tablenotemark{a} & HR2\tablenotemark{b}\\ 
\cline{4-5} \cline{6-7} \cline{8-9}\\
 & (h:m:s)& $(^{\circ}:\arcmin:\arcsec)$ &\colhead{Counts} & \colhead{S/N}
& \colhead{Counts} &
\colhead{S/N} &\colhead{Counts} & \colhead{S/N} & & }
\startdata

SSS-HR \\
M83-20 & 13:36:53.9&-29:48:48&   43.0 &   5.7 &   2.7 &   1.1 &   2.6 &  0.9 &  -0.9(-0.8)  & -0.9(-0.8)\\
M83-42 & 13:36:59.1&-29:53:36 &   30.5 &   5.3 &   0.0 &   0.0 &   0.0 &   0.0 &  -1.0(-0.9)  & -1.0(-0.9)\\
M83-50 &13:37:00.4&-29:50:54 &   105.1 &  10.0 &   0.8 &   0.3 &   0.0 &   0.0 &  -1.0(-0.9)  & -1.0(-1.0)\\
M83-54 & 13:37:01.1&-29:54:49 &  132.6 &  11.4 &   1.8 &   0.8 &   0.7 &   0.3 &  -1.0(-1.0)  & -1.0(-1.0)\\
M83-79 & 13:37:06.1&-29:52:32 &  108.6 &  10.3 &   1.8 &   0.9 &   0.0 &   0.0 &  -1.0(-0.9)  & -1.0(-1.0)\\
M83-81 & 13:37:06.5&-29:54:16&   15.7 &   3.7 &   0.0 &   0.0 &   0.0 &   0.0 &  -1.0(-0.8)  & -1.0(-0.8)\\
M83-88 & 13:37:07.4&-29:51:33 &   17.5 &   3.8 &   0.0 &   0.0 &   0.0 &   0.0 &  -1.0(-0.8)  & -1.0(-0.8)\\
M83-98 &13:37:12.8&-29:50:12&     35.9 &   5.8 &   2.1 &   1.0 &   1.6 &  0.7 &  -0.9(-0.8)  & -0.9(-0.8)\\
M83-111 & 13:36:59.5&-29:52:03&   78.2 &   8.4 &   5.9 &   2.2 &   2.0 &  0.9 &  -0.9(-0.8)  & -1.0(-0.9)\\
M83-128 & 13:36:57.7 & -29:53:53 & 29.6 & 5.2 & 1.6 & 0.7 & 0.0&0.0 &-0.9(-0.8) & -1.0(-0.8)\\
&&\\
SSS-$3\sigma$ \\
M83-1 & 13:36:40.8&-29:51:18&    25.2 &   4.7 &   2.9 &   1.3 &   1.6 &   0.6 &  -0.8(-0.7)  & -0.9(-0.8)\\
M83-7 &13:36:48.2 &-29:52:44 & 13.1 & 3.3 & 1.9 & 0.9 & 1.6 & 0.7 & -0.8(-0.6) & -0.8(-0.6)\\
M83-10 & 13:36:49.8&-29:52:17&   29.3 &  5.0 &   2.7 &   1.2 &   2.4 &   1.0 &  -0.8(-0.8)  & -0.8(-0.8)\\
M83-19 & 13:36:53.6&-29:56:00&   24.7 &   4.8 &   2.3 &   1.1 &   2.8 &   1.2 &  -0.8(-0.7)  & -0.8(-0.7)\\
M83-25 & 13:36:55.0&-29:52:39&   27.5 &   5.0 &   3.3 &   1.8 &   0.0 &   0.0 &  -0.8(-0.7)  & -1.0(-0.9)\\
M83-41 &13:36:58.7&-29:51:00 & 15.2 & 3.5 &   0.4 &   0.1 &   0.0 &   0.0 &  -1.0(-0.7)  & -1.0(-0.8)\\
M83-46 & 13:36:59.8&-29:55:25& 28.0 &   5.1 &   0.8 &   0.3 &   2.1 & 0.9 &  -1.0(-0.8)  & -0.9(-0.8)\\
M83-48 & 13:37:00.0&-29:54:17&   11.5 &   3.0 &   0.9 &  0.4 &   0.0 &   0.0 &  -0.9(-0.6)  & -1.0(-0.8)\\
M83-53 & 13:37:01.0&-29:50:56&   18.4 &   3.7 &   1.2 &  0.5 &   0.0 &   0.0 &  -0.9(-0.7)  & -1.0(-0.8)\\
M83-60 & 13:37:01.7&-29:51:13& 22.7 &   4.2 &   3.3 &   1.4 &   0.3 &   0.1 &  -0.7(-0.7)  & -1.0(-0.8)\\
M83-73 & 13:37:04.5&-29:51:08 &   14.0 &   3.1 &   0.0 &   0.0 &   0.0&   0.0 &  -1.0(-0.8)  & -1.0(-0.8)\\
M83-80 & 13:37:06.1&-29:54:44&   39.6 &   6.2 &   5.0 &   2.0 &   0.0 &   0.0 &  -0.8(-0.7)  & -1.0(-0.9)\\
M83-91 & 13:37:08.3&-29:51:26&   20.8 &   4.1 &   0.2 &   0.1 &   0.8 &   0.3 &  -1.0(-0.8)  & -0.9(-0.8)\\
M83-92 & 13:37:08.5&-29:51:35 &   12.8 &   3.0 &   0.0 &   0.0 &   0.0 &   0.0 &  -1.0(-1.0)  & -1.0(-1.0)\\
M83-93 & 13:37:11.8&-29:52:15&   45.0 &   6.5 &   4.6 &  1.8 &   1.3&   0.6 &  -0.8(-0.7)  & -0.9(-0.9)\\
M83-94 &13:37:12.1 &-29:50:56 & 14.2 & 3.4 & 0.6 & 0.2 & 0.6 & 1.0 &-0.9(-0.7)&-0.7(-0.6)\\
M83-97 &13:37:12.7&-29:52:00&    14.8 & 3.5 &   0.0 &   0.0 &   0.6 &   0.3 &  -1.0(-0.8)  & -0.9(-0.7)\\
M83-127 & 13:37:01.6&-29:52:01&   32.9 &   5.2 &   3.1 &   1.3 &   1.0 &   0.4 &  -0.8(-0.8)  & -0.9(-0.9)\\
&&\\
QSS-NOH\\
M83-12 & 13:36:50.9&-29:52:26&   12.0 &   2.8 &   3.7 &   1.6 &   0.0 &   0.0 &  -0.5(-0.5)  & -1.0(-0.8)\\
M83-22 &13:36:54.1&-29:52:09&    10.2 &   2.9 &   2.9 &   1.3 &   0.0 &   0.0 &  -0.6(-0.5)  & -1.0(-0.7)\\
M83-43 & 13:36:59.3&-29:55:08&    9.7 &   2.7 &   2.2 &   1.0 &   0.8 &   0.3 &  -0.6(-0.5)  & -0.9(-0.6)\\
M83-58 &13:37:01.6&-29:54:10&    10.1 &   2.8 &   2.9 &   1.3 &   0.0 &   0.0 &  -0.6(-0.5)  & -1.0(-0.7)\\
M83-85 & 13:37:07.0&-29:53:20&    6.6 &   2.2 &   1.0 &   0.4 &   0.0 &   0.0 &  -0.8(-0.5)  & -1.0(-0.7)\\
M83-86 & 13:37:07.1&-29:52:02&   15.8 &   3.7 &   3.2 &   1.4 &   0.0 &   0.0 &  -0.7(-0.6)  & -1.0(-0.8)\\
&&\\ 
QSS-SNOH\\
M83-15 & 13:36:52.8&-29:52:31&   20.0 &   3.9 &   7.5 &   2.5 &   0.0 &   0.0 &  -0.5(-0.4)  & -1.0(-0.8)\\
M83-24 & 13:36:55.0&-29:53:04&   21.2 &   4.0 &   7.8 &   2.6 &   0.0 &   0.0 &  -0.5(-0.4)  & -1.0(-0.9)\\

&&\\ 
QSS-HR$_1$\\
M83-11 & 13:36:50.8& -29:52:40& 20.3 & 3.8 & 7.5 & 2.5 & 1.7 & 0.7 & -0.5(-0.4) & -0.9(-0.7)\\
M83-13 & 13:36:51.1&-29:50:43 & 25.0 & 4.2 & 9.6 & 2.9 & 1.9 & 0.8 & -0.5(-0.4) & -0.9(-0.8)\\
M83-17 & 13:36:53.1&-29:53:25&   33.8 &   5.3 &   5.1 & 1.9 &   1.8 &  0.8 &  -0.7(-0.7)  & -0.9(-0.8)\\
M83-18 & 13:36:53.2&-29:52:42&   27.9 &   4.5 &  11.8 &   3.3 &   1.4 &  0.6 &  -0.4(-0.4)  & -0.9(-0.8)\\
M83-64 & 13:37:02.4&-29:51:26&   64.8 &   7.6 &  10.6 &   3.1 &   1.4 & 0.6 &  -0.7(-0.7)  & -1.0(-0.9)\\
M83-106 &13:37:17.4 &-29:51:54 & 13.8 & 3.5 & 5.0 & 2.0 & 1.7 & 0.7 &-0.5(-0.4) &-0.8(-0.6)\\
M83-114 &13:37:00.2&-29:52:06& 80.1 &   8.9 &  10.3 &   3.1 &   2.0 &   0.9 &  -0.8(-0.7)  & -0.9(-0.8)\\
M83-120 & 13:37:00.6&-29:52:04&  356.5 &  18.8 &  61.1 &   7.8 &   1.6 & 0.7 &  -0.7(-0.7)  & -1.0(-1.0)\\
&&\\
$QSS-3\sigma_1$\\
M83-89 & 13:37:07.5& -29:48:59& 11.5 & 2.2 & 2.0 & 0.8 & 1.4 &0.6 &-0.7(-0.6)&-0.8(-0.6)\\
M83-90 & 13:37:08.2&-29:49:16 & 10.3 & 2.3 & 2.0 & 0.8 &1.5 & 0.6 & -0.7(-0.5) & -0.8(-0.6)\\

&&\\
QSS-$\sigma$\\

M83-105 & 13:37:17.2&-29:51:53&   66.1 &   8.1 &  22.5 &   4.7 &   3.9 & 1.7 &  -0.5(-0.5)  & -0.9(-0.8)\\
M83-116 & 13:37:00.4&-29:51:59&  682.2 &  26.0 & 140.1 &  11.8 &  16.4 &   4.0 &  -0.7(-0.7)  & -1.0(-0.9)\\
M83-117 & 13:37:00.5&-29:51:56&  780.3 &  27.9 & 207.9 &  14.4 &  24.7 &   4.9 &  -0.6(-0.6)  & -0.9(-0.9)\\
M83-119 &13:37:00.6&-29:51:59&   526.6 &  22.9 & 110.4 &  10.5 &  10.0 &   3.0 &  -0.7(-0.6)  & -1.0(-1.0)\\

&&\\
QSS-FNOH\\
M83-66 & 13:37:03.0&-29:49:45&   11.5 &  2.5 &   4.0 &   1.6 &   0.0 &   0.0 &  -0.5(-0.4)  & -1.0(-0.8)\\
&&\\
QSS-MNOH\\
M83-35 & 13:36:57.8&-29:53:03&   68.6 &   8.1 &  13.1 & 3.5 &   0.0&   0.0 &  -0.7(-0.6)  & -1.0(-0.9)\\
M83-78 & 13:37:06.0&-29:55:14&   56.7 &   7.4 &  13.4 &   3.6 &   0.6 &   0.2 &  -0.6(-0.6)  & -1.0(-0.9)\\
M83-115 & 13:37:00.3&-29:52:05&   86.4 &   9.2 &  12.4 &   3.5 &   0.9 &   0.4 &  -0.8(-0.7)  & -1.0(-0.9)\\
\enddata

\tablenotetext{a}{(M-S)/(M+S); value in parentheses is the maximum of HR1.}
\tablenotetext{b}{(H-S)/(H+S); value in parentheses is the maximum of HR2.}

\end{deluxetable}

\begin{deluxetable}{lcccccccccc}
\tabletypesize{\scriptsize}
\tablewidth{0pt}
\tablecaption{SSS and QSS source list of M51}

\tablehead{
\multicolumn{1}{c}{Object}& R.A.& Dec. & \multicolumn{2}{c}{Soft} &
\multicolumn{2}{c}{Medium} & \multicolumn{2}{c}{Hard} & HR1\tablenotemark{a} & HR2\tablenotemark{b}\\ 
\cline{4-5} \cline{6-7} \cline{8-9}\\
 & (h:m:s)& $(^{\circ}:\arcmin:\arcsec)$ &\colhead{Counts} & \colhead{S/N}
& \colhead{Counts} &
\colhead{S/N} &\colhead{Counts} & \colhead{S/N} & & }
\startdata

SSS-HR\\
M51-12 & 13:29:43.3&+47:11:34&  272.7 &  16.5 &   5.2 &   2.1 &   1.0 &  0.4 &  -1.0(-0.9)  & -1.0(-1.0)\\
M51-42 & 13:29:55.4&+47:11:43&   14.7 &   3.6 &   0.0 &   0.0 &   0.0 &   0.0 &  -1.0(-0.8)  & -1.0(-0.8)\\
M51-58 & 13:30:02.3&+47:12:38&   20.4 &   4.5 &   0.0 &   0.0 &   0.0 &   0.0 &  -1.0(-0.9)  & -1.0(-0.9)\\
&&\\
SSS-$3\sigma$ \\
M51-14 & 13:29:44.0&+47:11:56&  7.9 &   2.6 &   0.0 &   0.0 &   0.9 &   0.4 &  -1.0(-0.7)  & -0.8(-0.5)\\
M51-17 & 13:29:45.9&+47:10:55&   11.9 &   3.3 &   0.0 &   0.0 &   0.0 &   0.0 &  -1.0(-0.8)  & -1.0(-0.8)\\
M51-22 & 13:29:50.1&+47:14:20&    8.7 &   2.8 &   0.0 &   0.0 &   0.0 &   0.0 &  -1.0(-0.7)  & -1.0(-0.7)\\
M51-27 &13:29:52.0 &+47:12:13 & 9.9 & 2.9 & 0.0 & 0.0 & 1.0 & 0.4 & -1.0(-0.7) & -0.8(-0.6)\\
M51-41 & 13:29:55.1&+47:10:42 &  8.1 &   2.6 &   0.0 &   0.0 &   0.8 &   0.4 &  -1.0(-0.7)  & -0.8(-0.5)\\
M51-43 & 13:29:55.4&+47:14:02&   10.6 &   3.0 &   0.0 &   0.0 &   0.9 &   0.4 &  -1.0(-0.8)  & -0.8(-0.6)\\
M51-45 & 13:29:56.0&+47:13:51&    7.7 &   2.4 &   1.0 &   0.4 &   0.7 &   0.3 &  -0.8(-0.5)  & -0.8(-0.5)\\
M51-59 & 13:30:04.1&+47:10:03&   10.2 &   3.1 &   2.1 &   1.0 &   0.8 &   0.3 &  -0.7(-0.5)  & -0.9(-0.6)\\
M51-62 & 13:30:05.7&+47:10:31&    8.2 &   2.7 &   0.9 &   0.4 &   0.0 &   0.0 &  -0.8(-0.5)  & -1.0(-0.7)\\
M51-64 &  13:30:06.0&+47:14:04&   9.8 &   3.0 &   1.9 &   0.9 &   0.5 &   0.2 &  -0.7(-0.5)  & -0.9(-0.6)\\
&&\\
QSS-NOH\\
M51-23 & 13:29:50.3&+47:13:22 &    4.7 &   1.9 &   1.0 &   0.4 &   0.0 &   0.0 &  -0.6(-0.4)  & -1.0(-0.6)\\
M51-70 & 13:30:20.9&+47:13:53&    5.2 &   1.9 &   3.4 &   1.4 &   0.9 &   0.3 &  -0.2(-0.2)  & -0.7(-0.4)\\
&&\\
QSS-SNOH\\
M51-6 & 13:29:38.6&+47:13:36 &  20.7 &   4.5 &   5.3 &   2.1 &   1.1 &  0.5 &  -0.6(-0.5)  & -0.9(-0.8)\\
M51-37 & 13:29:54.3&+47:11:21&   10.9 &   3.0 &   2.1 &   1.0 &   0.0 &   0.0 &  -0.7(-0.5)  & -1.0(-0.8)\\
M51-44 & 13:29:55.8&+47:11:44&   12.2 &   3.2 &   3.1 &   1.4 &   0.0 &   0.0 &  -0.6(-0.5)  & -1.0(-0.8)\\
&&\\
QSS-HR$_1$\\
M51-60 &13:30:04.5 &+47:10:31 & 11.2 & 3.3 & 6.1 & 2.3 & 2.0 & 0.9 & -0.3(-0.2) & -0.7(-0.6) \\

&&\\
QSS-$\sigma$\\

M51-10 & 13:29:39.9&+47:12:36&  201.9 &  14.2 &  70.4 &   8.4 &   9.2 &   2.9 &  -0.5(-0.5)  & -0.9(-0.9)\\
&&\\
QSS-FNOH\\
M51-5 &13:29:38.1&+47:16:12&    51.7 &   7.1 &  10.8 &   3.2 &   0.0 &   0.0 &  -0.7(-0.6)  & -1.0(-0.9)\\
M51-32 & 13:29:53.5&+47:11:26&    9.8 &   2.6 &   3.1 &   1.4 &   0.9 &   0.4 &  -0.5(-0.4)  & -0.8(-0.6)\\
M51-57 & 13:30:02.1&+47:12:38&    9.1 &   2.9 &   4.1 &   1.8 &   0.0 &   0.0 &  -0.4(-0.3)  & -1.0(-0.7)\\

\enddata

\tablenotetext{a}{(M-S)/(M+S); value in parentheses is the maximum of HR1.}
\tablenotetext{b}{(H-S)/(H+S); value in parentheses is the maximum of HR2.}

\end{deluxetable}

\begin{deluxetable}{lcccccccccc}
\tabletypesize{\scriptsize}
\tablewidth{0pt}
\tablecaption{SSS and QSS source list of NGC4697}

\tablehead{
\multicolumn{1}{c}{Object}& R.A.& Dec. & \multicolumn{2}{c}{Soft} &
\multicolumn{2}{c}{Medium} & \multicolumn{2}{c}{Hard} & HR1\tablenotemark{a} & HR2\tablenotemark{b}\\ 
\cline{4-5} \cline{6-7} \cline{8-9}\\
 & (h:m:s)& $(^{\circ}:\arcmin:\arcsec)$ &\colhead{Counts} & \colhead{S/N}
& \colhead{Counts} &
\colhead{S/N} &\colhead{Counts} & \colhead{S/N} & & }
\startdata

SSS-HR\\
NGC4697-16 & 12:48:37.1&-05:47:58 &   63.0 &   7.9 &   0.0 &   0.0 &   0.0 &   0.0 &  -1.0(-1.0)  & -1.0(-0.9)\\
NGC4697-19 & 12:48:34.5&-05:47:49&   73.7 &   8.5 &   0.9 &   0.4 &   0.0 &   0.0 &  -1.0(-0.9)  & -1.0(-1.0)\\
NGC4697-52 & 12:48:41.2&-05:48:19&   58.6 &   7.5 &   0.0 &   0.0 &   0.7 &   0.3 &  -1.0(-0.9)  & -1.0(-0.9)\\

&&\\
SSS-3$\sigma$\\
NGC4697-91 & 12:48:38.9 & -05:47:31 &12.7 & 3.5 & 0.0 & 0.0 & 0.0 & 0.0 & -1.0(-0.8) & -1.0(-0.8)\\
&&\\
QSS-NOH\\
NGC4697-73 & 12:48:24.1&-05:48:16&   17.3 &   4.0 &   3.6 &   1.6 &   0.2 &   0.0 &  -0.7(-0.6)  & -1.0(-0.8)\\
NGC4697-86 & 12:48:19.6&-05:47:33&    8.2 &   2.6 &   3.1 &   1.4 &   0.4 &   0.2 &  -0.5(-0.4)  & -0.9(-0.6)\\
&&\\
QSS-SNOH\\
NGC4697-2 & 12:48:36.0&-05:48:03 &  15.7 &   3.9 &   8.4 &   2.8 &   0.9 &   0.4 &  -0.3(-0.3)  & -0.9(-0.7)\\
NGC4697-9 & 12:48:35.4&-05:47:54 &   40.2 &   6.2 &   7.0 &   2.5 &   0.0 &   0.0 &  -0.7(-0.7)  & -1.0(-0.9)\\
NGC4697-10 & 12:48:36.5&-05:48:01&   11.5 &   3.3 &   6.3 &   2.4 &   0.0 &   0.0 &  -0.3(-0.3)  & -1.0(-0.8)\\
NGC4697-17 & 12:48:37.0&-05:47:53&   13.6 &   3.5 &   6.2 &   2.3 &   0.5 &   0.2 &  -0.4(-0.3)  & -0.9(-0.7)\\
&&\\
QSS-HR$_1$\\

NGC4697-13 & 12:48:36.2&-05:48:15&   12.5 &   3.1 &   4.1 &   1.7 &1.3 &   0.5 &  -0.5(-0.4)  & -0.8(-0.6)\\
NGC4697-29 & 12:48:33.3& -05:48:02& 15.3 & 3.7 & 6.8 & 2.4 & 1.8 & 0.8 & -0.4(-0.4) & -0.8(-0.7)\\
NGC4697-49 &12:48:31.0 &-05:48:28 & 11.8 & 3.1 & 7.0 & 2.5 & 2.3 & 1.0 & -0.3(-0.2) &-0.7(-0.5)\\
NGC4697-50 & 12:48:40.8&-05:48:22&   13.1 &   3.3 &   2.8 &   1.2 &   1.4 &   0.6 &  -0.7(-0.6)  & -0.8(-0.6)\\
NGC4697-59 &12:48:32.9 &-05:46:14 & 11.4 & 3.3 & 7.9 & 2.7 & 1.8 & 0.8 & -0.2(-0.2) &-0.7(-0.6)\\
NGC4697-66 & 12:48:41.4&-05:46:02 & 16.4 & 3.9 & 12.1 & 3.4 & 2.3 & 1.0 & -0.2(-0.1)&-0.8(-0.6)\\
NGC4697-74 &12:48:28.7 &-05:50:25 &23.4 & 4.4 & 9.1 & 2.9 & 9.1 & 0.9 &-0.4(-0.4) &-0.8(-0.7)\\
&&\\
QSS-FNOH\\
NGC4697-68 & 12:48:25.5&-05:48:08&  7.6 &   2.5 &   6.7 &   2.4 &   0.5 &   0.2 &  0.0(0.0)  & -0.9(-0.6)\\
&&\\
QSS-$\sigma$\\
NGC4697-64 & 12:48:34.4&-05:50:14&   51.2 &   6.9 &  23.6 &   4.8 &   3.2 &   1.2 &  -0.4(-0.4)  & -0.9(-0.8)\\

\enddata

\tablenotetext{a}{(M-S)/(M+S); value in parentheses is the maximum of HR1.}
\tablenotetext{b}{(H-S)/(H+S); value in parentheses is the maximum of HR2.}

\end{deluxetable}

\begin{deluxetable}{lccccccc}
\tabletypesize{\scriptsize}
\tablewidth{0pt}
\tablecaption{Spectral Fits to the Brightest SSSs and QSSs}

\tablehead{
ID& $N_H$& kT\tablenotemark{a} & kT$_{RS}$\tablenotemark{b} &
$\alpha$\,\tablenotemark{c}
&$\chi^2_{\nu}$/dof& Flux\,\tablenotemark{d} & $L_X$\,\tablenotemark{e}\\
  &  ($10^{21}$cm$^{-2}$)& (eV) & (keV) & & &}
\startdata
SSS-HR & & & &\\
M101-104 & $1.16^{+1.08}_{-0.44}$ & $67^{+8}_{-14}$ &&& 1.39/9
&2.11&0.74, 1.13\\ 
M101-51 & $2.14^{+0.83}_{+0.19}$ & $50^{+3}_{-12}$ &&& 0.57/7
&6.19&2.16, 3.32\\
M101-18 & 1 (fixed) & $71^{+18}_{-19}$ & & & 0.96/13 & 1.02 & 0.35, 0.55\\
M101-102 & $1.54^{+0.19}_{-0.31}$& $91^{+7}_{-4}$&
$0.74^{+0.04}_{-0.05}$&&1.18/75&82.67&28.8, 44.4\\
M83-50& $0.24^{+0.74}_{-0.24}$ & $66^{+13}_{-24}$ &&& 1.24/5 & 1.14 &
0.28, 0.30\\
M83-54& $0.95^{+1.38}_{-0.69}$ & $71^{+10}_{-16}$ &&& 0.66/7 & 2 &
0.50, 0.53\\
M83-79 & $2.1^{+3.74}_{-1.53}$ & $86^{+27}_{-30}$ &&& 0.33/5 & 4 & 1,
1.1\\
M51-12 & $0.5^{+0.92}_{-0.46}$ & $102^{+12}_{-16}$ &&& 1.12/15 &
6.17&4.38, 5.20\\
NGC4697 (3 sources)& $0.50^{+0.31}_{-0.37}$ & $81^{+11}_{-7}$ &&& 0.66/6
& 1.89&3.10, 12.28\\
&&&&\\
SSS-$3\sigma$&&&&\\
M83 (18 sources)& $1.83^{+1.66}_{-0.88}$ & $124^{+21}_{-24}$&&& 1.05/17 &
9.01 & 2.25, 2.38\\
&&&&\\
QSS-MNOH&&&&\\
M101-114 & $4.4^{+2.33}_{-1.52}$& $124^{+27}_{-25}$ &&&0.91/16& 9.0&
3.14, 4.83\\
M101-116 & 1 (fixed) & $158^{+27}_{-21}$ & && 1.34/10 & 0.45 & 0.05, 0.08\\
M83-35& $2.59^{+7.6}_{-1.52}$& $138^{+37}_{-76}$ &&& 1.03/11 & 9.47&
2.37, 2.50\\
&&&&\\
QSS-FNOH&&&&\\
M51 (3 sources)& $4.46^{+5.39}_{-4.55}$& $110^{+60}_{-110}$ &&& 0.97/4 &
35.4&25.1, 29.9\\
&&&&\\
QSS-SNOH&&&&\\
M101 (3 sources)& $1.20^{+1.09}_{-0.48}$ & & & $3.20^{+1.32}_{-0.40}$ &
 1.06/10&1.15& 0.40, 0.62\\
&&&&\\
QSS-$\sigma$&&&&\\
M101-53 & $6.72^{+1.13}_{-2.45}$ & &$0.64^{+0.10}_{-0.07}$ && 1.18/7 &
2.01 & 0.7, 10.8\\
M51-10 & 0.2 (fixed) & $222^{+26}_{-23}$ &&& 1.16/13 & 4.69& 3.33, 3.96\\
M83-105 & $2.76^{+3.42}_{-1.63}$ & $165^{+45}_{-55}$ &&& 1.13/5&1.36&
0.34, 0.36\\
M83-116 & $2.14^{+2.94}_{-1.18}$ & $146^{+33}_{-45}$ &&& 1.55/25 &
6.17& 1.54, 1.63\\
M83-117\tablenotemark{f} & $4.82^{+0.81}_{-1.31}$ & $54^{+18}_{-15}$ &
$0.51\pm0.05$~\tablenotemark{g} &&1.31/19 & 75.6 & 18.9, 20\\
M83-119 & $2.77^{+3.26}_{-1.43}$ & $132^{+27}_{-37}$&&& 1.24/20 & 6.5
& 1.62, 1.72\\
&&&&\\
QSS-NOH&&&&\\
M83 (6 sources)& $0.74^{+3.66}_{-0.74}$ & $172^{+74}_{-74}$ &&& 0.57/5 &
0.95& 0.24, 0.25\\
&&&&\\
QSS-HR$_1$&&&& \\
M83-120\tablenotemark{f} & 1 (fixed) & $56^{+3}_{-4}$ &
$0.68^{+0.06}_{-0.05}$~\tablenotemark{g} & & 0.62/14 & 4.74 & 1.18, 1.25\\

\enddata
\tablecomments{All quoted uncertainties are 90\% confidence. The ACISABS
model was applied to correct for the degradation of ACIS (see text).}
\tablenotetext{a}{Blackbody temperature.}
\tablenotetext{b}{Raymond-Smith temperature.}
\tablenotetext{c}{Power-law photon index.}
\tablenotetext{d}{Unabsorbed flux ($\times10^{-14}$ erg
cm$^{-2}$ s$^{-1}$) in 0.3--7 keV.}
\tablenotetext{e}{Luminosity ($\times10^{38}$ erg s$^{-1}$) in 0.3--7
keV. The upper and lower limits quoted here are based on the
farthest and nearest distances, respectively,
shown in Table 2.}
\tablenotetext{f}{Background (diffuse emission) not subtracted.} 
\tablenotetext{g}{MEKAL plasma temperature.}
\end{deluxetable}

\begin{deluxetable}{lccccc}
\tablewidth{0pt}
\tablecaption{Central Concentration of VSSs}

\tablehead{
Galaxy& $f_{VSS}$ & $f_{all}$ & $f_{area}$ & ${\cal F}_{VSS}=f_{VSS}/f_{area}$ &
${\cal F}_{all}=f_{all}/f_{area}$}
\startdata
M101 & 0.09 & 0.08 & 0.006 & 15 & 13\\
M83 & 0.20 & 0.18 & 0.004 & 50 & 45\\
M51 & 0.05 & 0.04 & 0.001 & 50 & 40\\
NGC4697 & 0.36 & 0.14 & 0.001 & 360 & 140\\
\enddata
\tablecomments{$f_{VSS}$: Fraction of VSSs in the central 1 kpc; $f_{all}$
Fraction of all X-ray sources in the central 1 kpc; $f_{area}$: fraction
of the galaxy's total surface area comprised by a $1$ kpc radius disk.}
\end{deluxetable}

\clearpage

\begin{figure}
{\rotatebox{-90}{\psfig{file=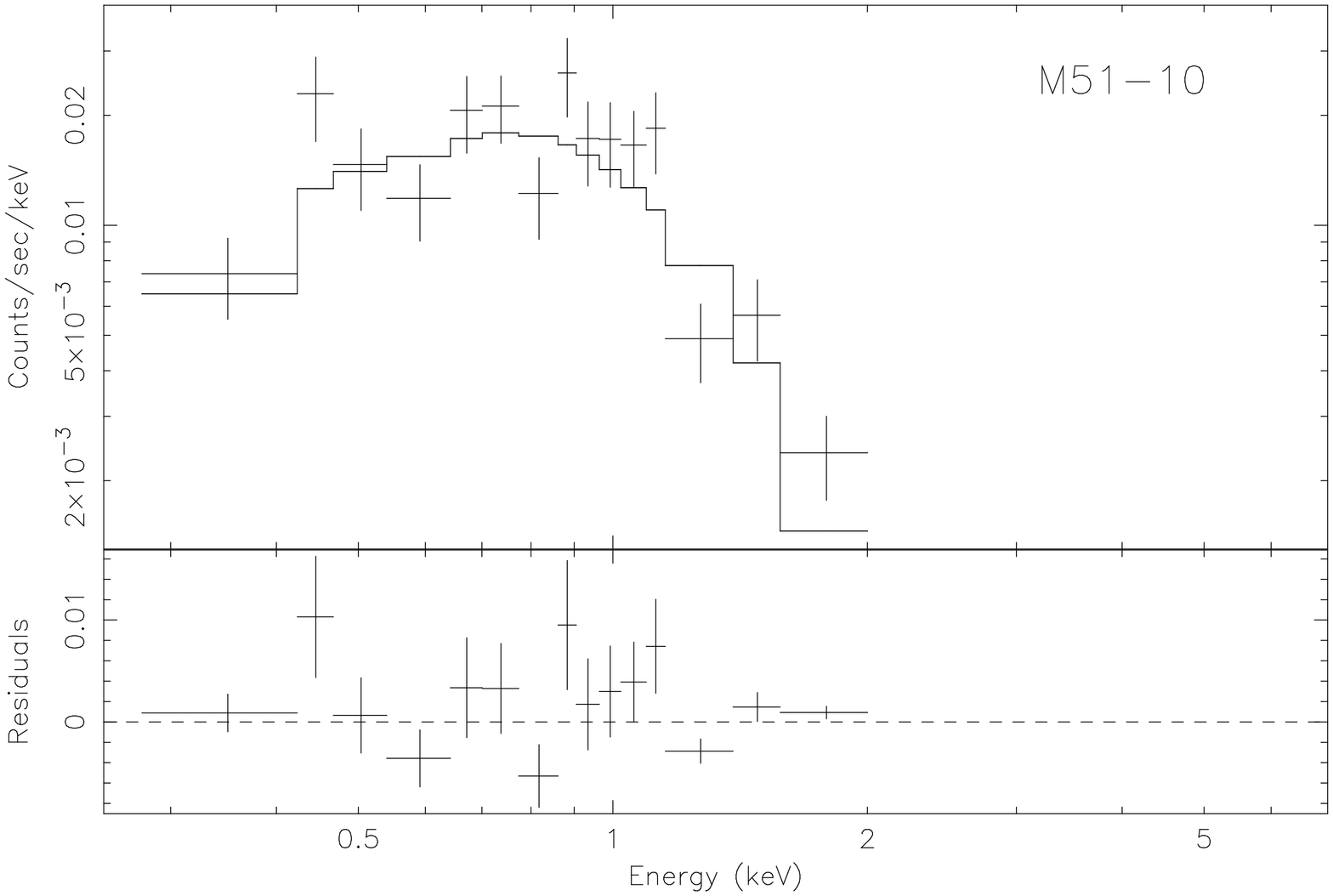,height=3.5in}}}
{\rotatebox{-90}{\psfig{file=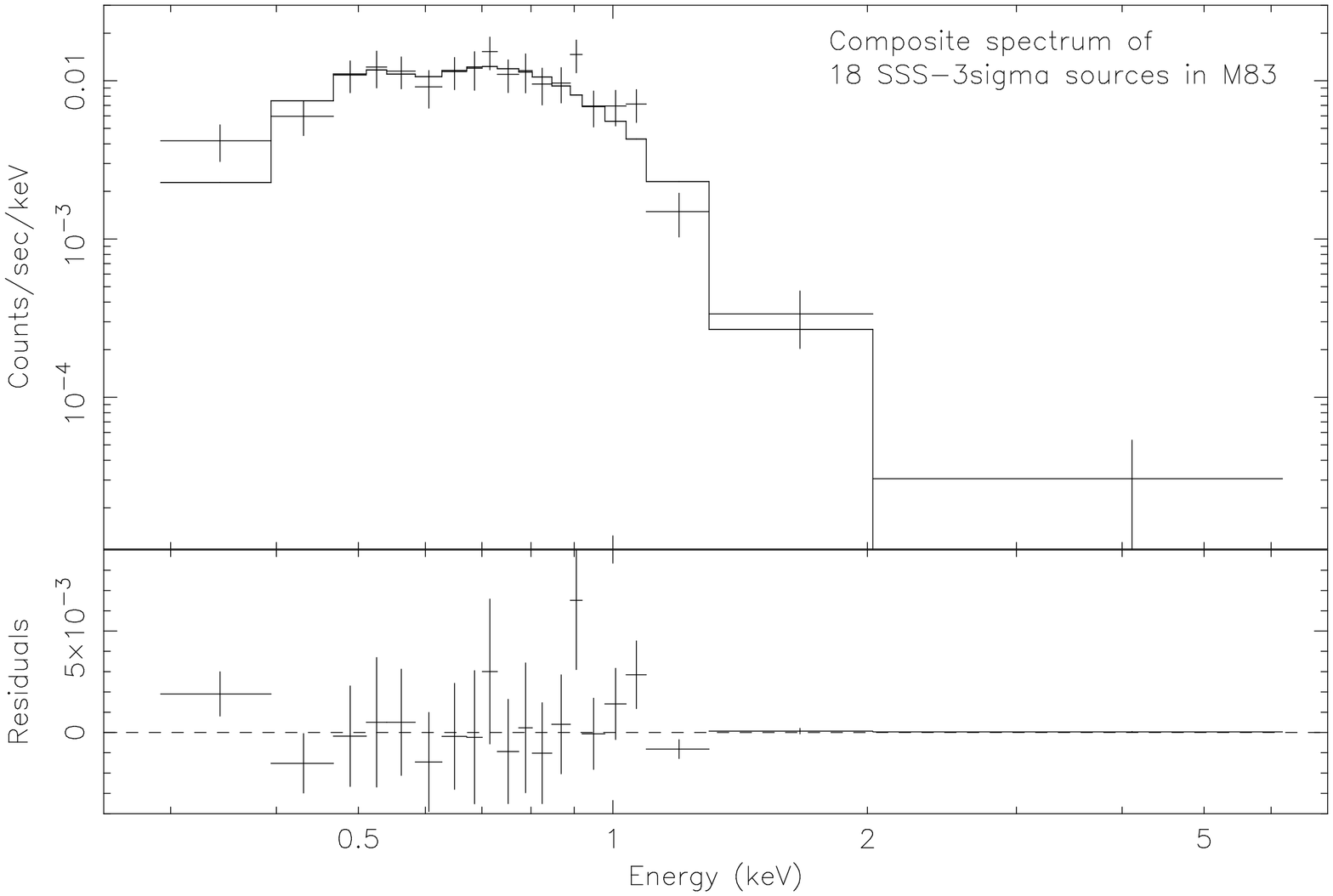,height=3.5in}}}
\caption{Left: Blackbody spectral fit to QSS-$\sigma$ M51-10 ($N_H=2\times10^{20}$cm$^{-2}$, $kT=222$ eV). Right: Composite spectrum ($N_H=1.8\times10^{21}$cm$^{-2}$, $kT=124$ eV) of 18 SSS-$3\sigma$ sources in M83. Fits to other SSSs are shown in \rd\ and Kong (2003b).
}
\end{figure}

\begin{figure}
\psfig{file=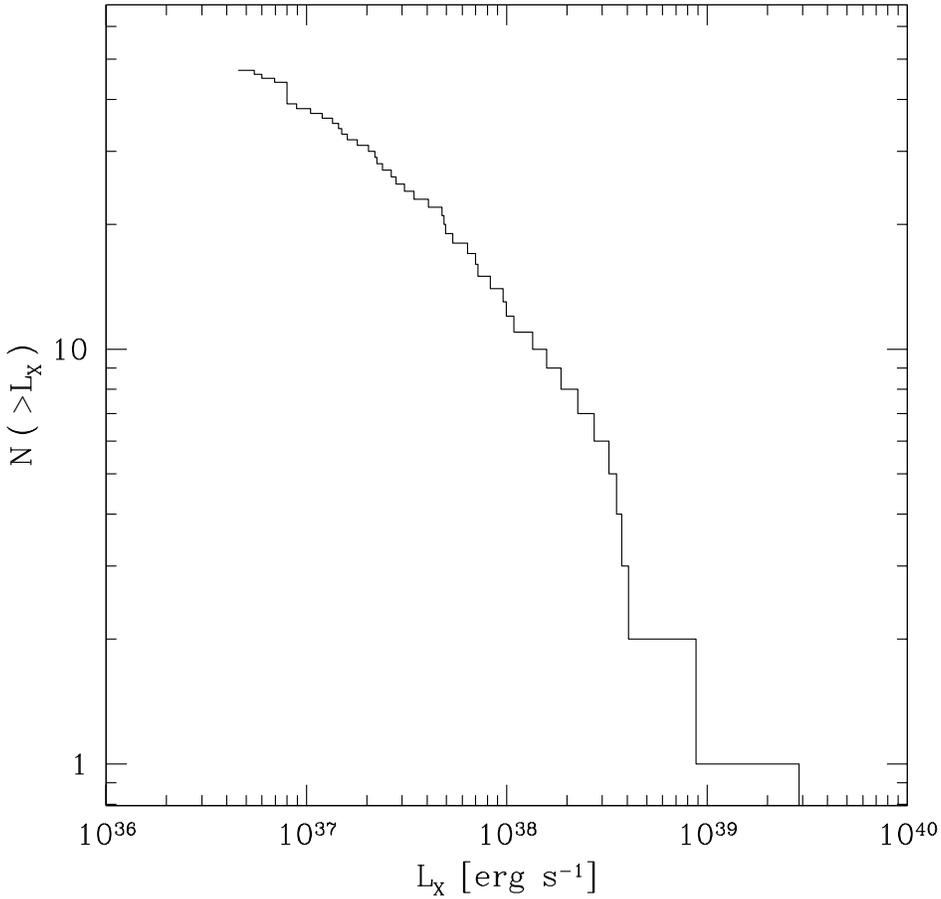,height=5in}
\caption{Cumulative luminosity function of all VSSs
for which we have spectral fits.
(See Table 6 and \S 2.2.1.) The luminosities are based on the lower limit of the distance estimation for the galaxy listed in Table 1.
The brightest source 
is M101-102, with a high (low) luminosity of
$4.4 \times 10^{39}$ ergs s$^{-1}$ ($2.9 \times 10^{39}$ ergs s$^{-1}$).}
\end{figure}

\begin{figure}
\psfig{file=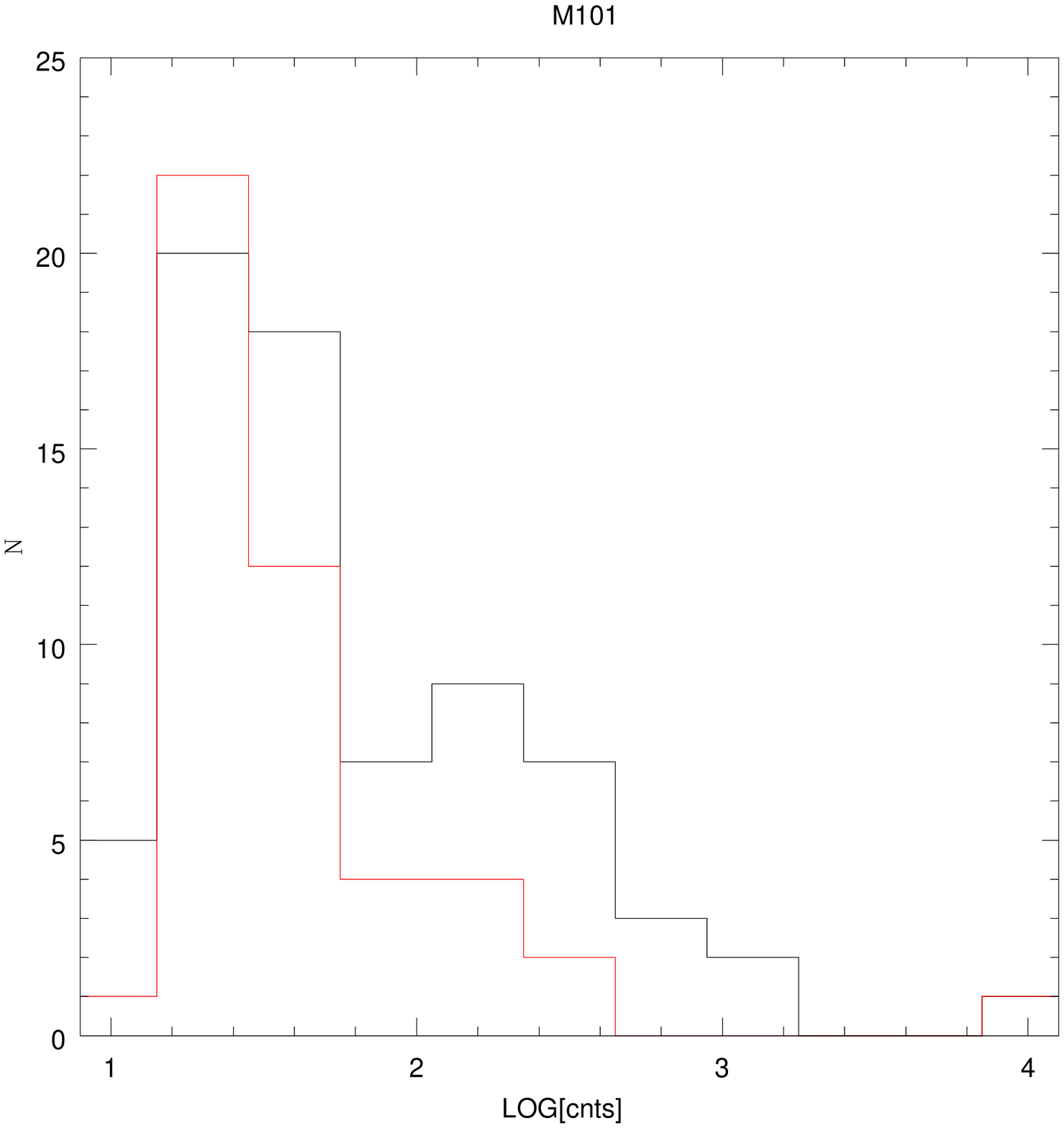,width=3.6in}
\psfig{file=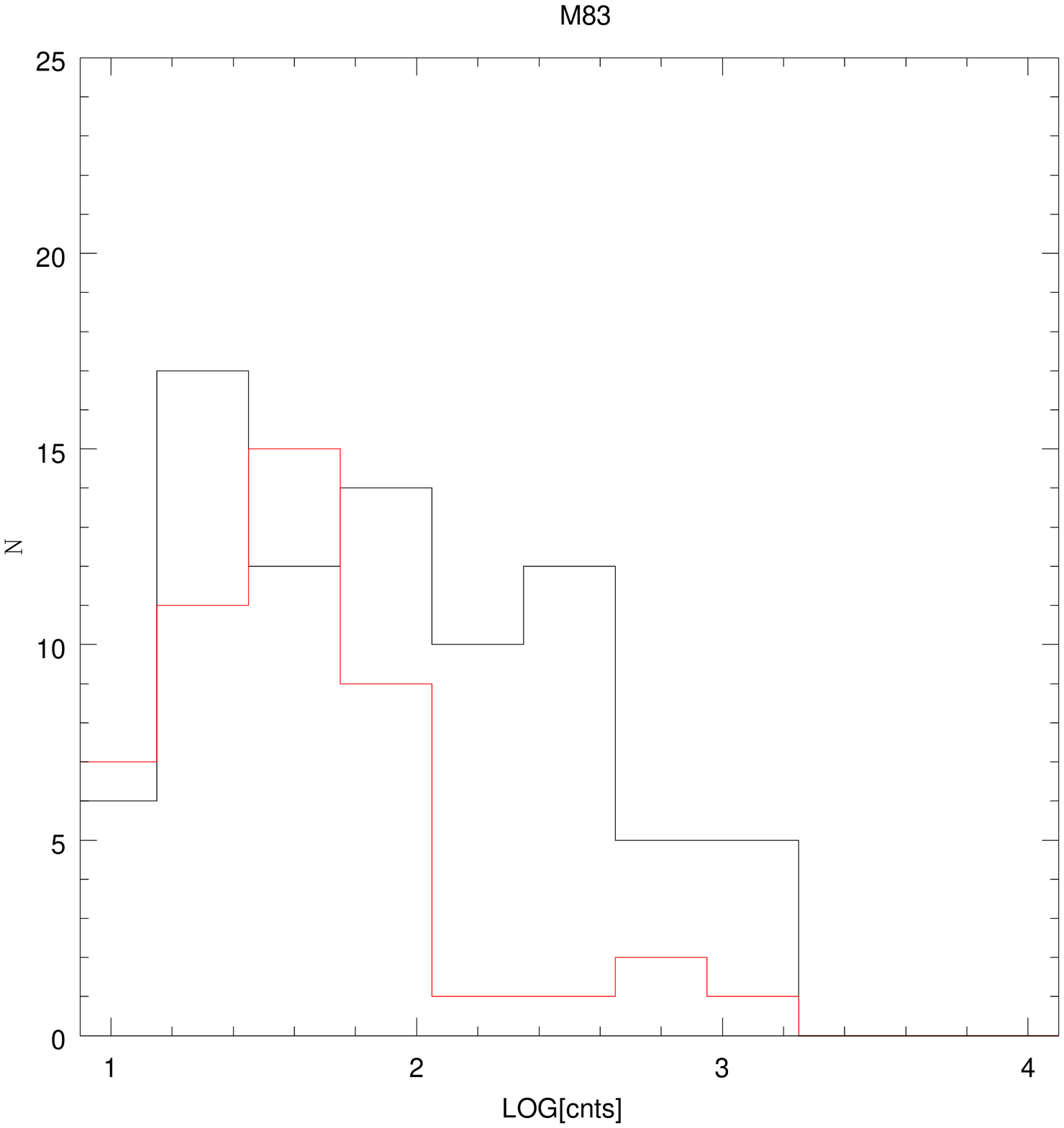,width=3.6in}
\psfig{file=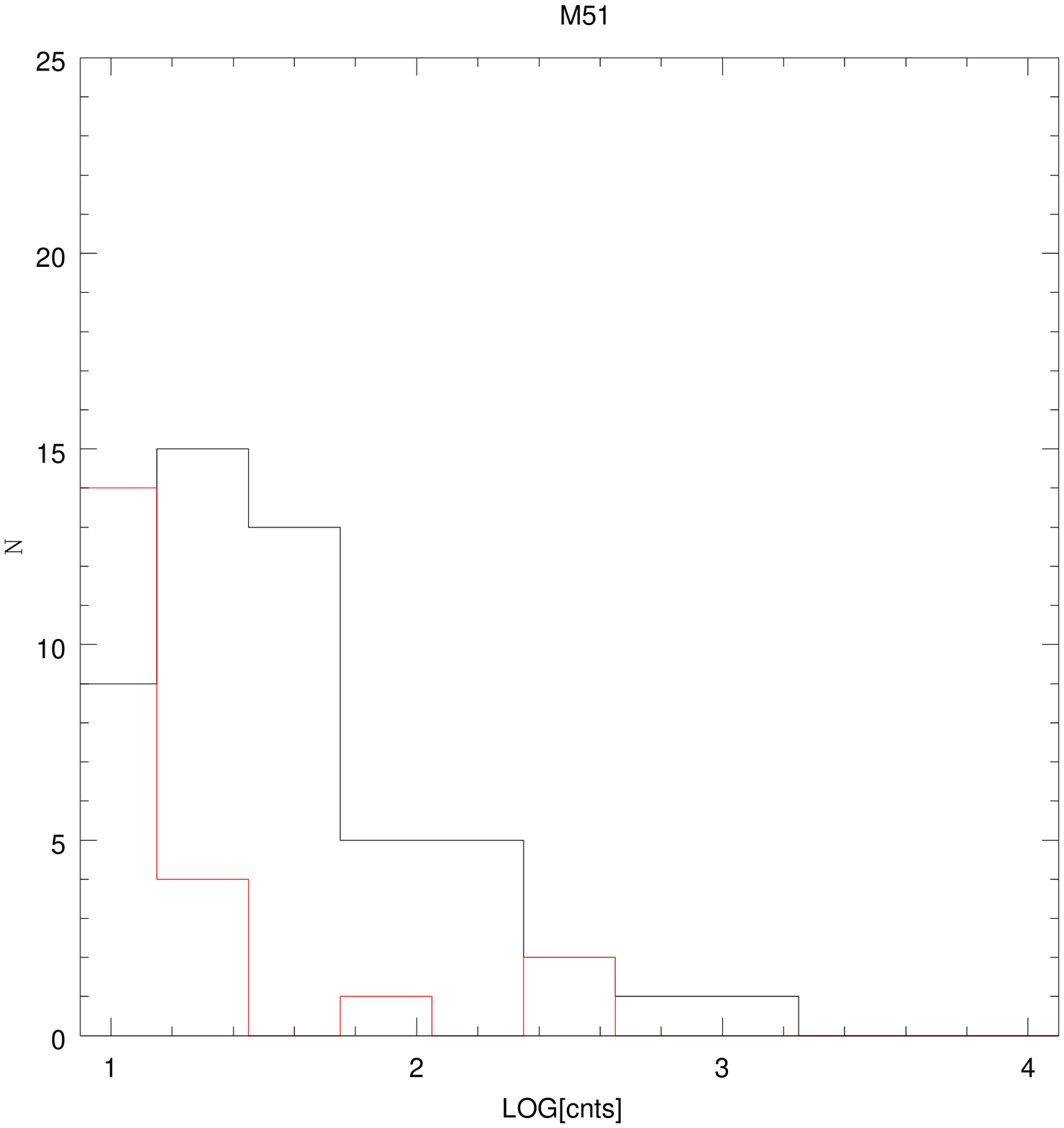,width=3.6in}
\psfig{file=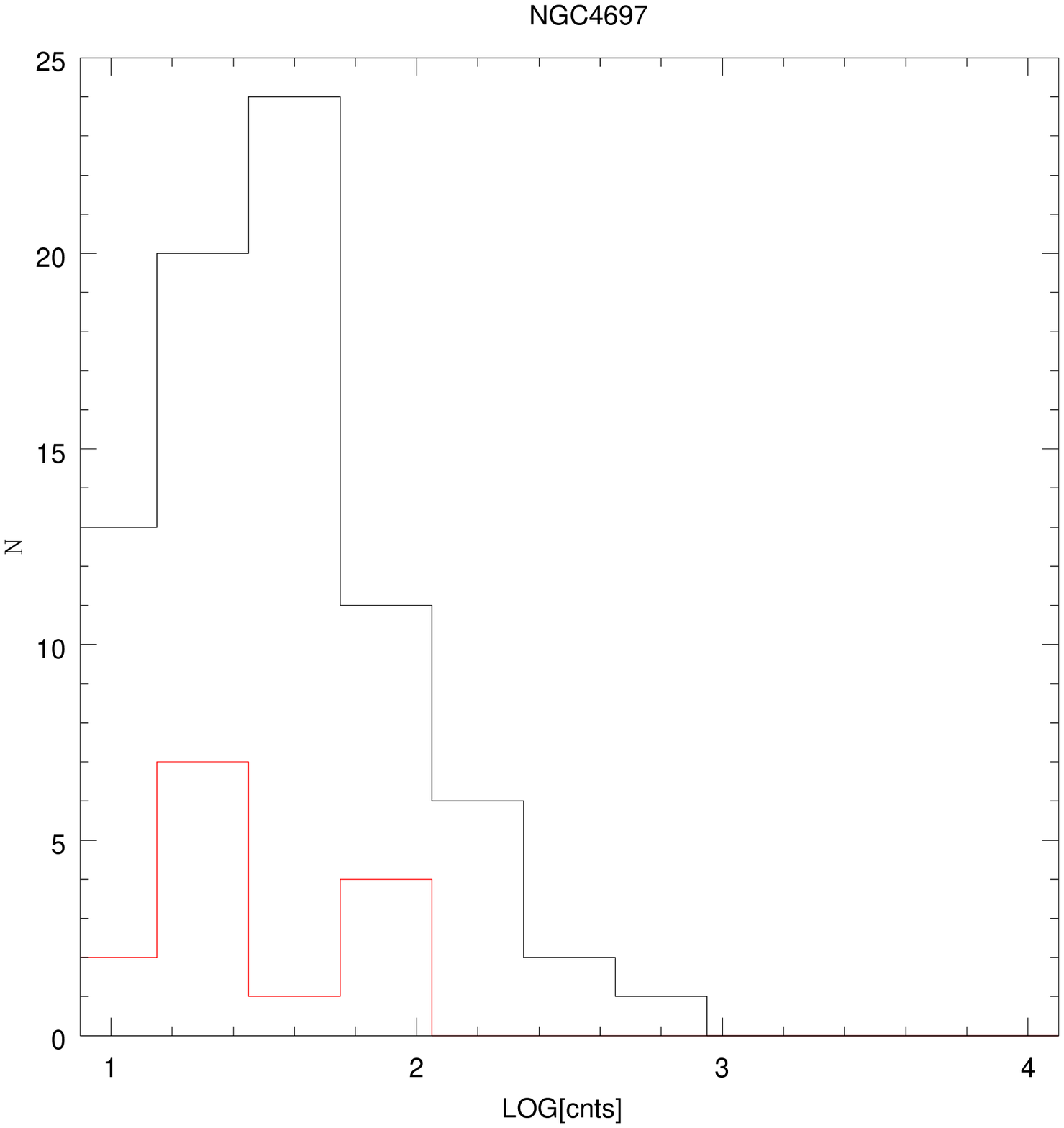,width=3.6in}
\caption{Histograms of total counts for each galaxy. 
Red: VSSs; black: non-VSSs. See \S 5.3.2.}
\end{figure}

\begin{figure}
\epsfig{file=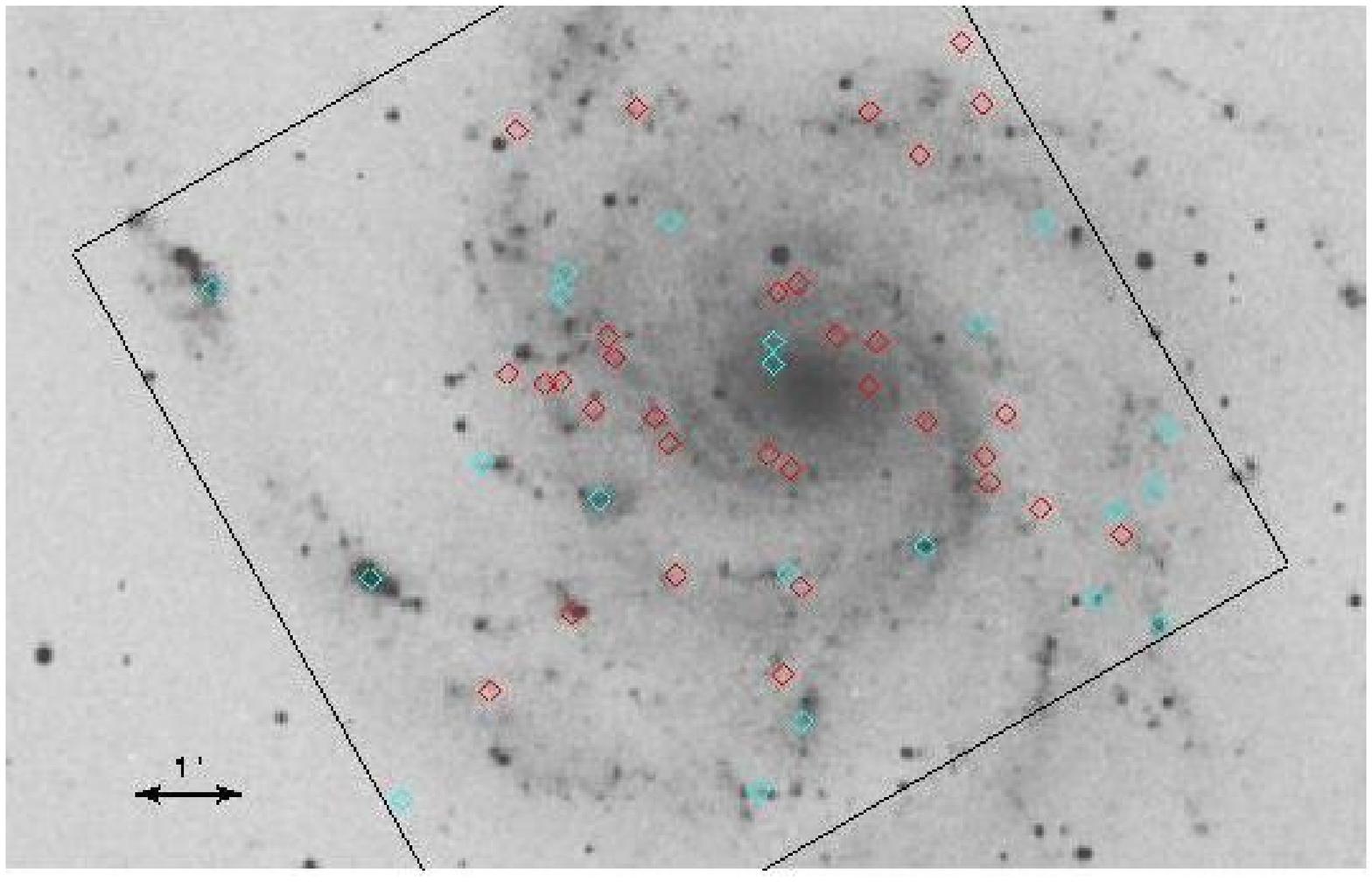,width=3.6in}
\epsfig{file=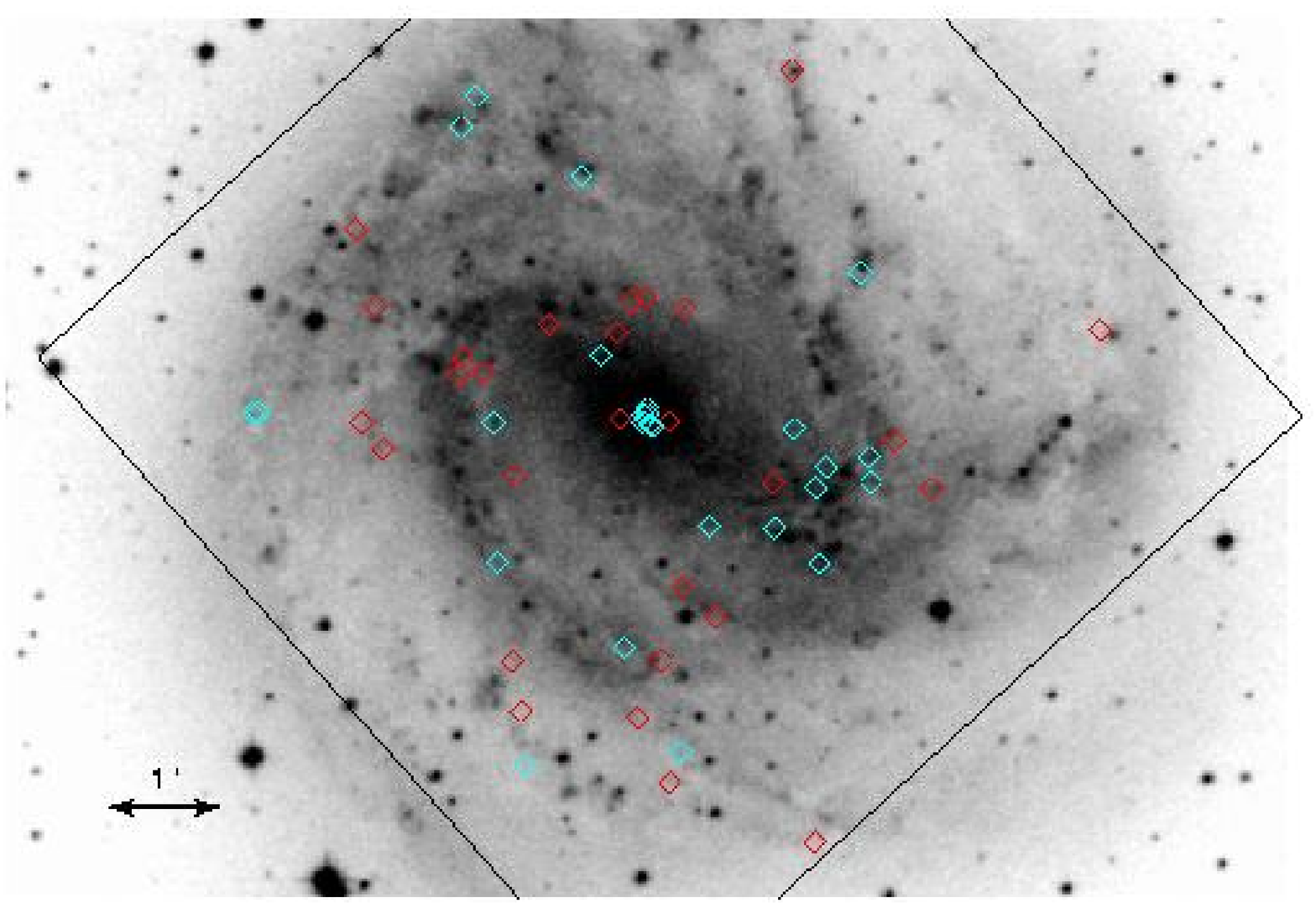,width=3.6in}
\epsfig{file=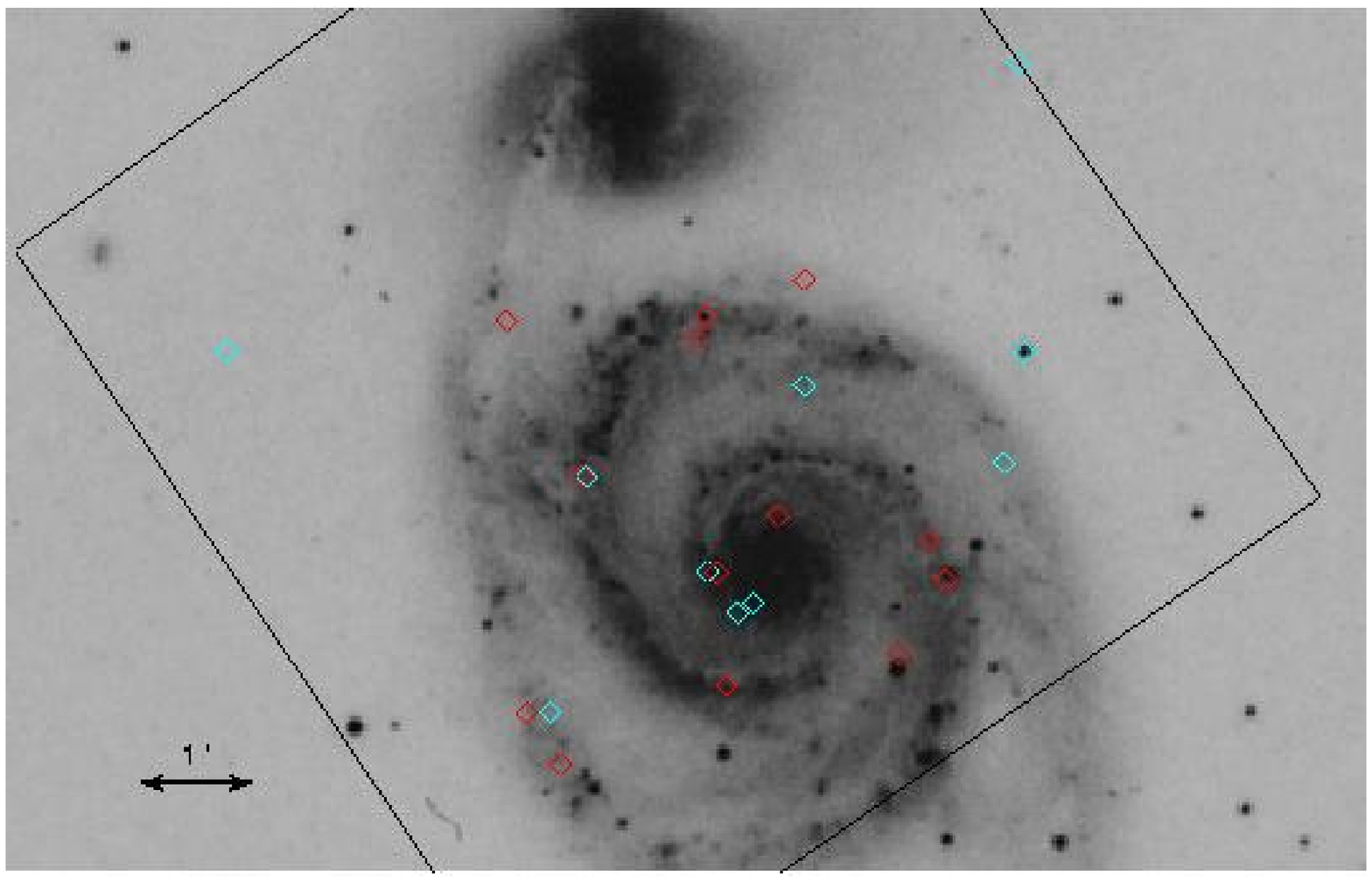,width=3.6in}
\epsfig{file=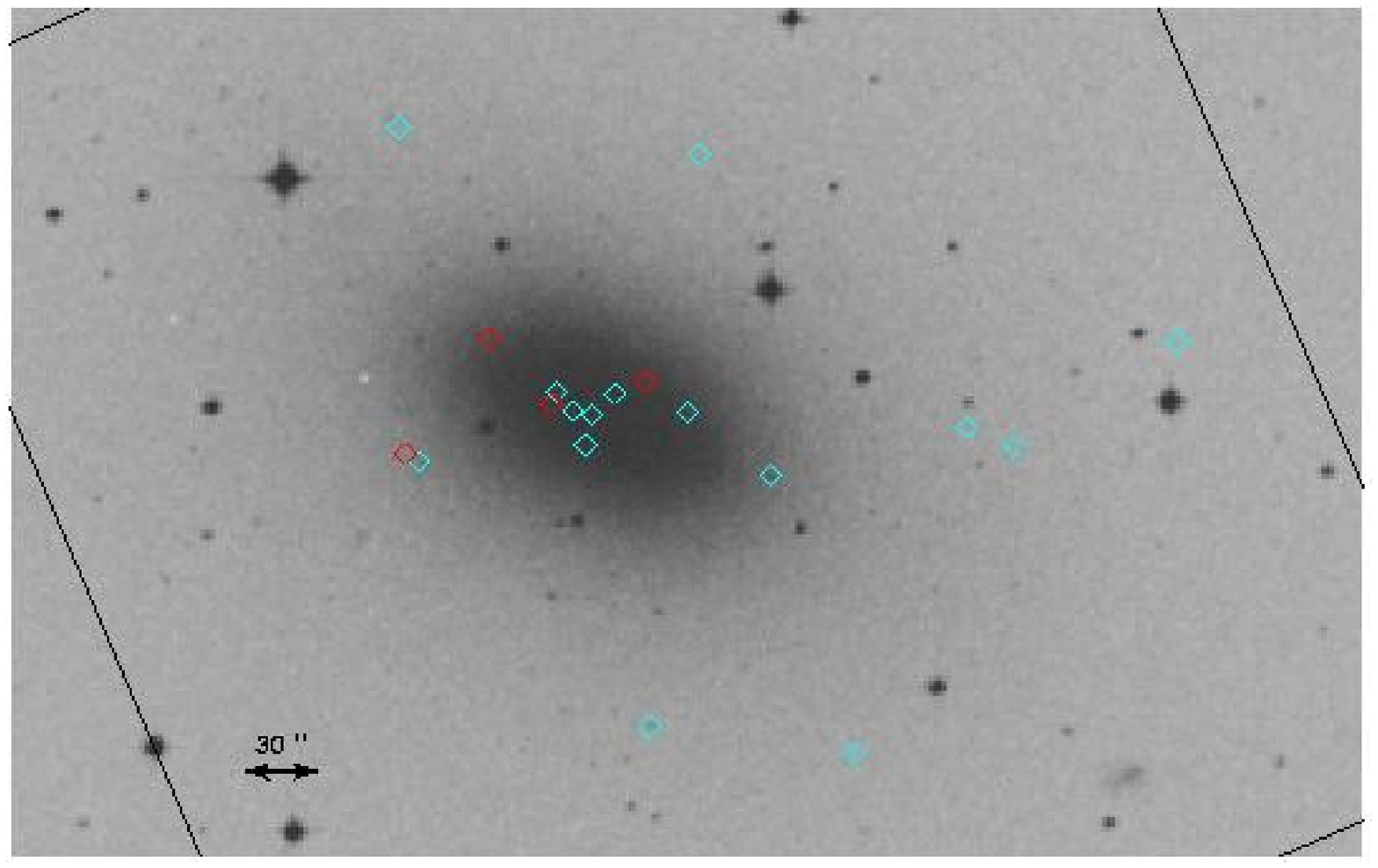,width=3.6in}
\caption{{\it Chandra} field-of-view (ACIS-S3; black line) overlaid on
the Digital Sky Survey images of M101 (upper left), 
M83 (upper right), M51 (lower left) and NGC4697 (lower right). Also
shown in the figures are the positions of all very soft X-ray sources (SSSs: red;
QSSs: cyan).}
\end{figure}

\begin{figure}
\epsfig{file=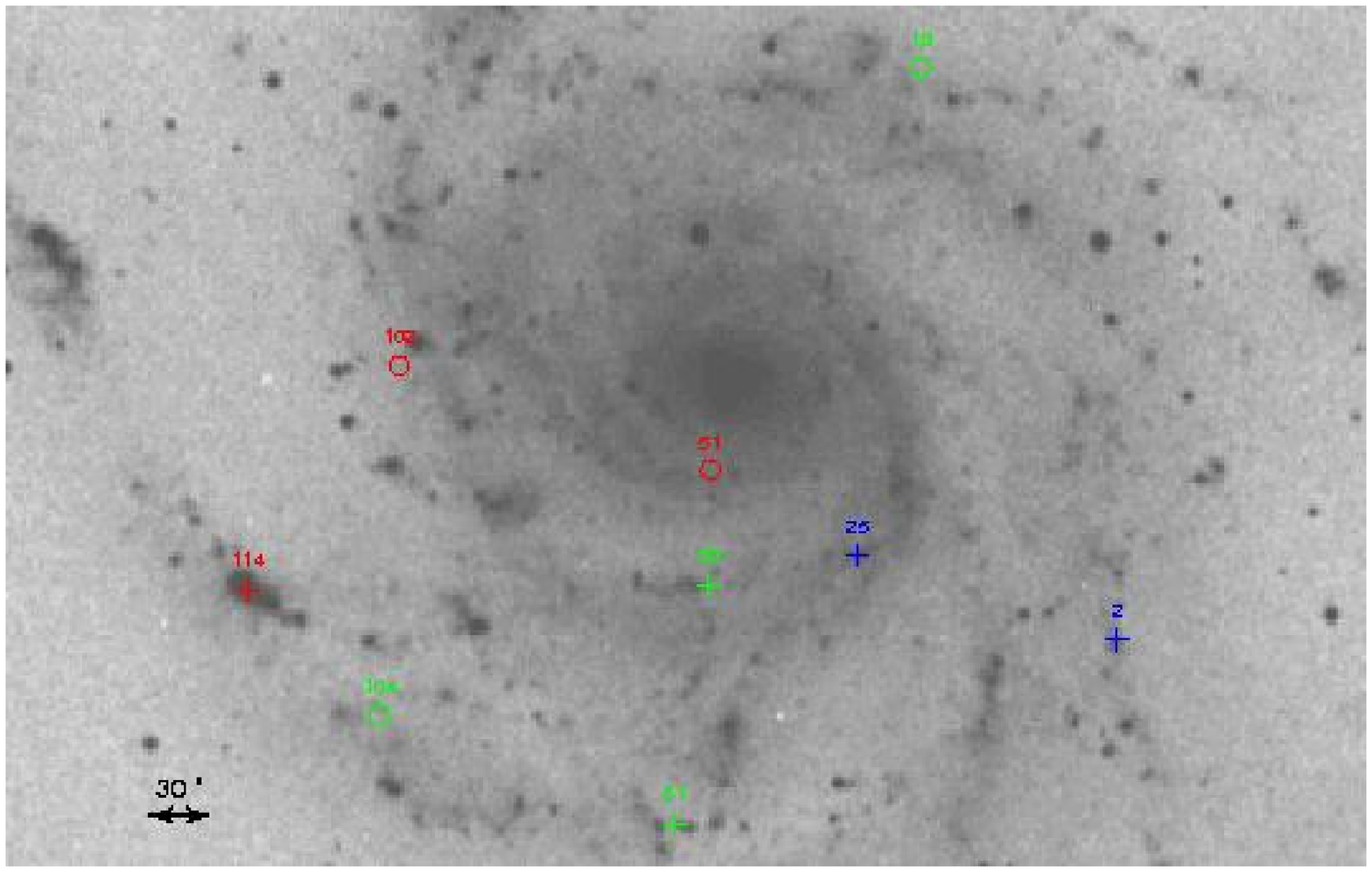,width=3.6in}
\epsfig{file=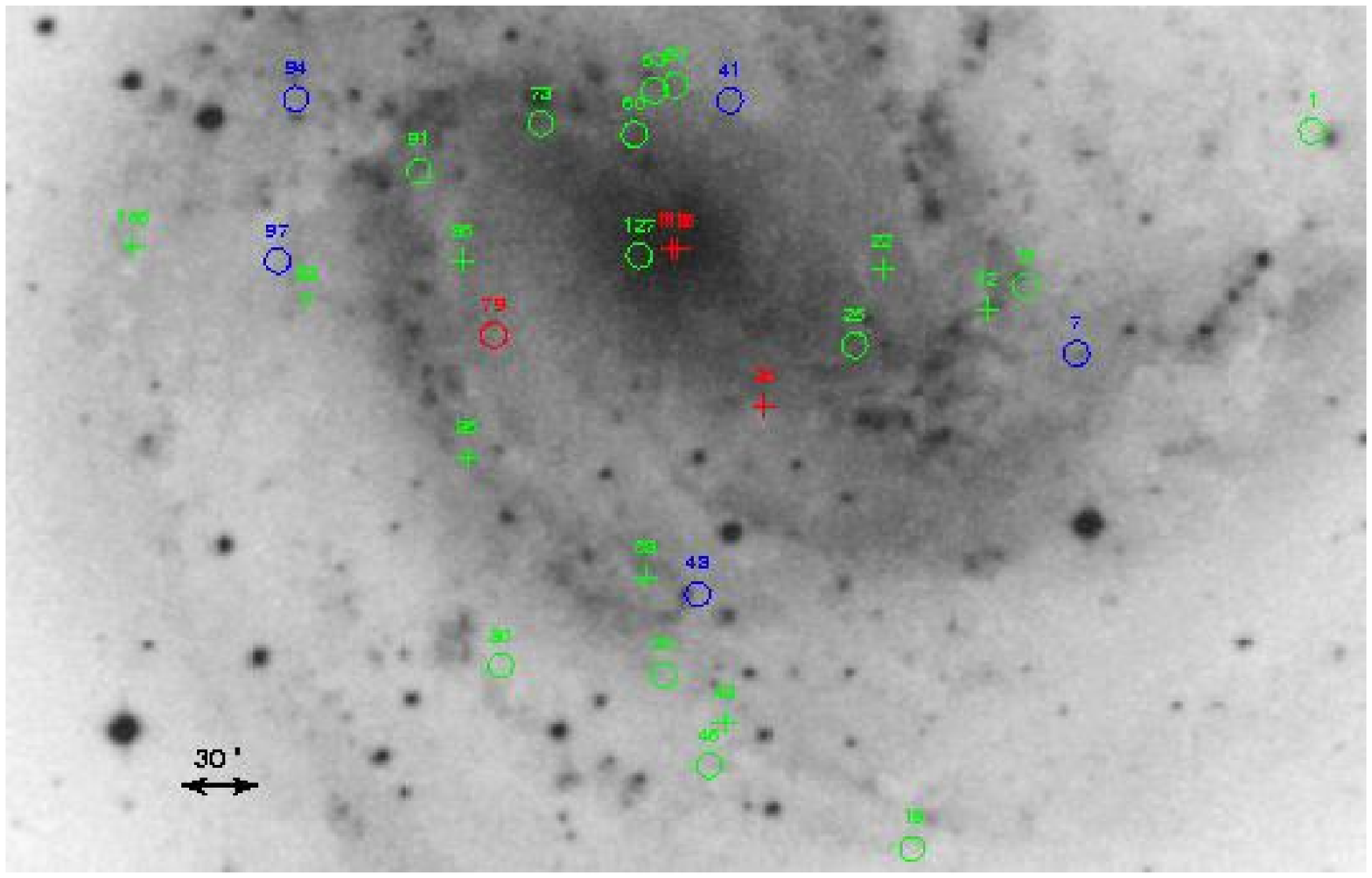,width=3.6in}
\epsfig{file=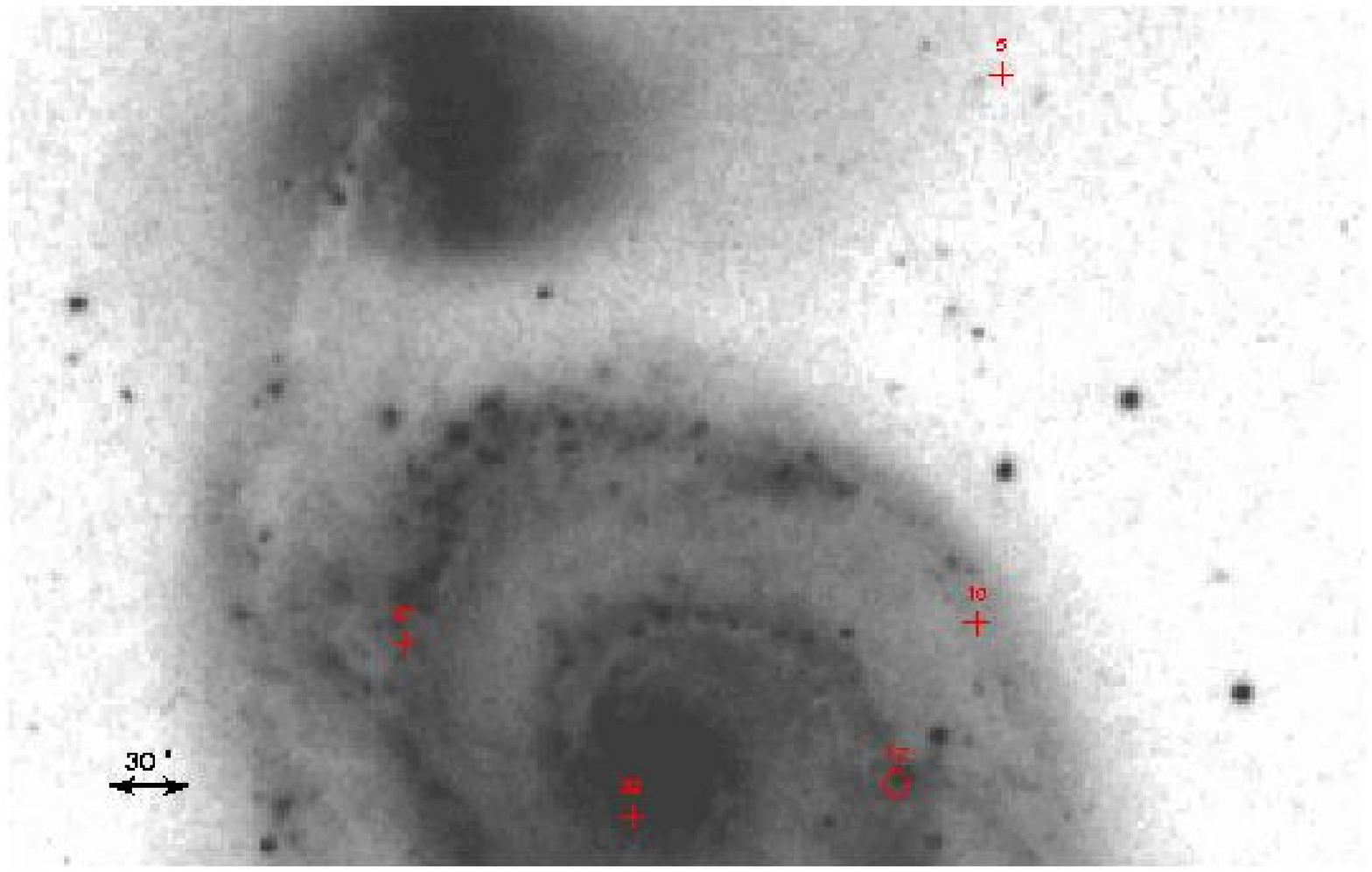,width=3.6in}
\caption{Very soft sources which provided enough photons to allow spectral fits.
The galaxies are arranged as in Figure 10. 
Red: luminosity above $10^{38}$ erg s$^{-1}$;
green: luminosity between $10^{37}$ erg s$^{-1}$ and $10^{38}$ erg
s$^{-1}$; blue: luminosity between 
$10^{36}$ erg s$^{-1}$ and $10^{37}$ erg
s$^{-1}$) in M101, M83 and M51. Circles represent SSSs while crosses are QSSs. 
The number is the source number listed in Table 2, 3, and 4.}
\end{figure}

\begin{figure}
\epsfig{file=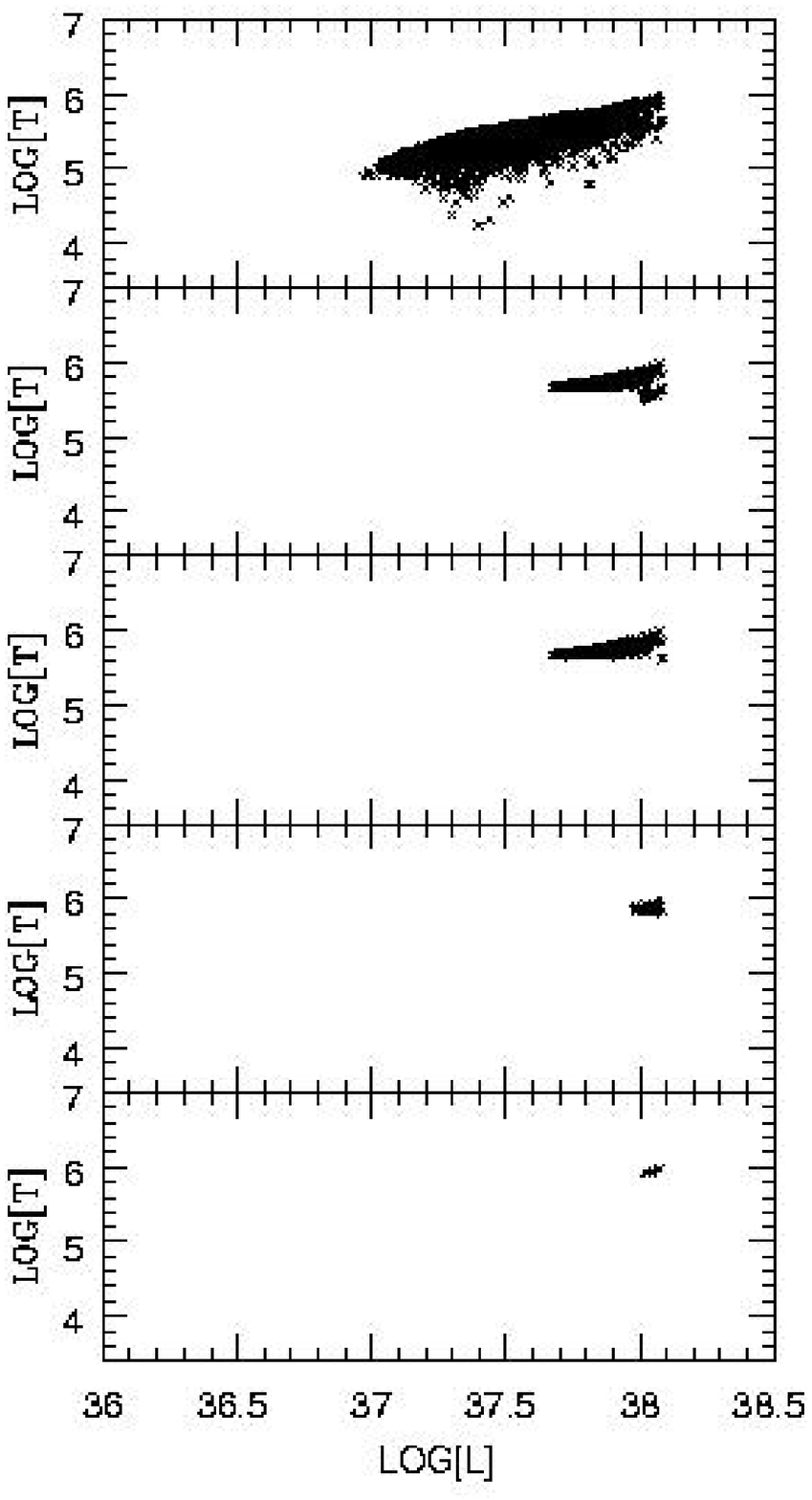,width=3in}
\caption{$Log[T]$ vs $Log[L]$ for (top panel) all ``classical" SSSs 
from an RDS distribution (see \S 6); those members of the 
distribution detected and selected as SSSs in
M101, M83, M51, and NGC 4697 (successive lower panels).}
\end{figure}


\begin{references}


\reference{}Chiang, E.\& Rappaport, S. 1996, ApJ, 469, 255


\reference{} Di Stefano, R., Greiner, J., Murray, S.,
 Garcia, M.  2001, ApJ,. 51L, 37


\reference{} Di Stefano, R.~\& Kong, A.~K.~H.\ 2003a, \apj, 592, 884

\reference{} Di\,Stefano, R.~\& Kong, A.~K.~H.\  2003b, \apj, submitted (astro-ph/0311374)

\reference{} Di\,Stefano, R. et al. 2003, \apj, submitted (astro-ph/0306440)

\reference{} Di\, Stefano, R. 1996, in
Supersoft X-Ray Sources, 
Proceedings of the International Workshop Held in Garching, Germany,
28 February - 1 March 1996.
Lecture Notes in Physics, Vol. 472, edited by Jochen Greiner.
Springer-Verlag, Berlin Heidelberg New York, 1996., p.193
 

\reference{} Di Stefano, R., Paerels, F.,
\& Rappaport, S.
1995, ApJ, 450, 705 

\reference{} Di\, Stefano, R., Rappaport, S. 1994, ApJ, 437, 733

\reference{} Fabbiano, G., Kim, D.-W., Trinchieri, G. 1994, ApJ, 428, 555


\reference{} Faber, S.M., Wegner, G., Burstein, D., Davies, R.L., Dressler, A., Lynden-Bell, D.,
 Terlevich, R. J. 1989, ApJS, 71, 173

\reference{}Feldmeier, J.J., Ciardullo, R., Jacoby, G.H. 1997, ApJ, 479, 231

\reference{}Freedman, W.L., et al. 2001, ApJ, 553, 47

\reference{}Greiner J. 2000 New Astronomy 5, 137.


\reference{}Greiner, J., Tovmassian, G.H.,
 \rd, R., Prestwich, A.,
 Gonzalez-Riestra, R., Szentasko, L.,
 \& Chevarria, C. 1999,  A\&A, 343, 183

\reference{}Greiner, J. \& van Teeseling  1998, A\&A, 339L, 21



\reference{}Hagiwara, Y, Henkel, C, Menten, K.M., Nakai, N. 2001 ApJ, 560, L37



\reference{}Ho, L. C., Filippenko, A. C., \& Sargent, W. L. W. 1997, ApJS, 112, 315

\reference{} Ho, L.C. 2002, ApJ, 564, 120

\reference{} Kenyon, S.J. 1986, The Symbiotic Stars, CUP

\reference{} Karachentsev, I.D., et al. 2002, A\&A, 385, 21

\reference{}Kong, A.K.H., Garcia, M.R., Primini, F.A., \& Murray, S.S. 2002, ApJ, 580, L125 

\reference{} Kong, A.~K.~H., Sjouwerman, L.~O., Williams, B.~F., Garcia, M.~R., \& 
Dickel, J.~R.\ 2003, \apjl, 590, L21

\reference{}Kong, A.K.H., \& Di\,Stefano, R. 2003, \apjl, 590, L13

\reference{} Kong, A.K.H., \& Di\,Stefano, R. 2004, ATel, 222



\reference{}Long, K.S., Helfand, D.J., \& Grabelsky, D.A. 1981, ApJ, 248, 925 

\reference{} Matonick, D.M., \& Fesen, R.A. 1997, ApJS, 112, 49



\reference{}Moody, J. W., Roming, P.W.A., Joner, M.D., Hintz, E.G., Geisler, D.,
Durrell, P.R., Scowen, P.A., Jee, R.O. 1995, AJ, 110, 2088 


\reference{} Mukai, K., Pence, W.D., Snowden S.L., \& Kuntz, K.D. 2002, ApJ, 582, 184

\reference{} Nedialkov, P., Orio, 
M., Birkle, K., Conselice, C., Della Valle, M., Greiner, J., Magnier, E., 
\& Tikhonov, N.~A.\ 2002, \aap, 389, 439


\reference{} Patterson, J., 
Thorstensen, J.~R., Fried, R., Skillman, D.~R., Cook, L.~M., \& Jensen, L.\ 
2001, \pasp, 113, 72

\reference{} Pence, W. D., Snowden, S. L.,
 Mukai, K., \& Kuntz, K. D. 2001, ApJ, 561, 189 



\reference{} Rappaport, S., Chiang, E.,
 Kallman, T., \& Malina, R. 1994, ApJ, 431, 237


\reference{} Rappaport, S., Di\, Stefano, R.,
 Smith, J.D.  1994, ApJ, 426, 692



\reference{} Sarazin, Craig L., Irwin, Jimmy A.,
 Bregman, Joel N,  2001, ApJ, 556, 533

\reference{} Scowen P.A., Dufour R.J., \& Hester J.J. 1992, AJ, 104, 92



\reference{} Soria, R. \& Wu, K. 2002, A\&A, 384, 99

\reference{} Soria, R. \& Wu, K. 2003, A\&A, 410, 53

\reference{} Swartz, D.A.,
 Ghosh, K.K.,
 Suleimanov, V.,
 Tennant, A.F., \& Wu, K. 2002, ApJ, 574, 382  

\reference{} Terashima, Y. \& Wilson, A.S. 2003, ApJ, 601, 735

\reference{} Thatte, N., Tecza, M., \& Genzel, R. 2000, A\&A, 364, L47

\reference{} Tonry, J.L., Dressler, A., Blakeslee, J.P., Ajhar, E.A., Fletcher,
A.B., Luppino, G.A., Metzger, M.R., Moore, C.B. 2001, ApJ, 546, 681

\reference{} Tully, R.B. 1988, in Nearby Galaxies Catalog, Cambridge and New York, Cambridge University Press,


\reference{} Vanlandingham, K.~M., Starrfield, S., 
Shore, S.~N., \& Sonneborn, G.\ 1999, \mnras, 308, 577 



\end{references}
\end{document}